\documentclass[
aip,
amsmath,amssymb,
preprint
]{revtex4-1}

\usepackage{graphicx}% Include figure files
\usepackage{dcolumn}% Align table columns on decimal point
\usepackage{bm}% bold math
\usepackage[utf8]{inputenc}
\usepackage[T1]{fontenc}
\usepackage{mathptmx}
\usepackage{etoolbox}
\usepackage{xcolor}
\usepackage{subfigure}
\makeatletter
\def\@email#1#2{%
    \endgroup
    \patchcmd{\titleblock@produce}
    {\frontmatter@RRAPformat}
    {\frontmatter@RRAPformat{\produce@RRAP{*#1\href{mailto:#2}{#2}}}\frontmatter@RRAPformat}
    {}{}
}%
\makeatother
\begin{document}
    
    \preprint{AIP/123-QED}
    
    \title[Generalized Taylor dispersion for translationally invariant microfluidic systems]{Generalized Taylor dispersion for translationally invariant microfluidic systems}
    % Force line breaks with \\
    \author{A. Alexandre}
    \affiliation{Univ. Bordeaux, CNRS, Laboratoire Ondes et Mati\`ere d'Aquitaine (LOMA), UMR 5798, F-33405 Talence, France}%Lines break automatically or can be forced with \\
    \author{T. Gu\'erin}
    \affiliation{Univ. Bordeaux, CNRS, Laboratoire Ondes et Mati\`ere d'Aquitaine (LOMA), UMR 5798, F-33405 Talence, France}%Lines break automatically or can be forced with \\ 
    \author{D.S. Dean}
    \email{david.dean@u-bordeaux,fr}
    \affiliation{Univ. Bordeaux, CNRS, Laboratoire Ondes et Mati\`ere d'Aquitaine (LOMA), UMR 5798, F-33405 Talence, France}

    \date{\today}% It is always \today, today,
    
    \begin{abstract}
        We consider Taylor dispersion for tracer particles in micro-fluidic planar channels with strong confinement. In this context, the channel walls modify the local diffusivity tensor and also interactions between the tracer particles and the walls become important. We provide a simple and general formula for the effective diffusion constant along the channel as well as the first non-trivial finite time correction for arbitrary flows along the channel, arbitrary interaction potentials with the walls and arbitrary expressions for the diffusion tensor. The formula are in particular amenable to a straightforward numerical implementation, rendering them extremely useful for comparison with experiments. We present a number of applications, notably for systems which have parabolically varying diffusivity profiles, to systems with attractive interactions with the walls as well
        as electroosmotic flows between plates with differing surface charges within the Debye-H\"uckel approximation.
    \end{abstract}
    
    \maketitle
    
    \section{Introduction}
    \label{sec:intro}
    
    Taylor dispersion is a classic example  \cite{tay53,ari56,fra89} of how  the spreading of tracer particles is influenced by the properties of the surrounding medium, in particular by  the presence of hydrodynamic  flows. It  has been extensively studied in the context of macroscopic fluid mechanics and a number of analytical methods have been adapted for its study. The first method, as implemented by Taylor \cite{tay53} and then Aris \cite{ari56}, is based on the averaging of the Fokker-Planck equation describing the full advection-diffusion problem. Subsequently, methods based on the direct computation of the moments of the particle dispersion along the channel were developed
    \cite{bar83,ved14, adro2019a, adro2019b}. These methods have the advantage that for simple systems the full temporal dependence of the moments can be obtained. Brenner and coworkers applied macro-transport theory \cite{fra89,bre93} to give a very general and powerful method to compute long-time dispersion properties. A similarly general statistical mechanics based theory, based on Kubo-type formulas, has also been formulated \cite{gue15a,gue15b}.  The center manifold approximation has also been used to compute the dominant late time behavior of dispersion moments and in some simple cases can be pushed to compute the late time moments of very high orders \cite{mer90,mer94,bal95}.  Taylor dispersion has also been studied for systems where the tracer particles can be absorbed at the surface of a uniform channel, this absorption can be either irreversible or reversible and the absorbed tracer can diffuse on the channel surface or be considered immobile \cite{bre93,bal95,lev12,ber13}.
    
    In microfluidic systems, a number of new questions, which are largely irrelevant for large scale systems, arise when analyzing Taylor dispersion.  First, the presence of channel walls modifies the diffusion tensor of the tracer particle when the tracer size becomes of the same order as its distance from the walls. While analytic, though complicated, expressions for the microscopic local diffusion constant parallel and perpendicular to a single wall exist \cite{hap91}, the case where two walls are present can only be treated  approximately or numerically \cite{ose27,gan80,duf01}. Clearly, in highly confined systems, the variation of the diffusion constant in the direction perpendicular to the channel will have a strong effect on the Taylor dispersion. In addition, interactions (notably electrostatic and van der Waals) with the walls will become important. While this is taken into account via the absorption models discussed above for macroscopic systems, the interaction with the wall must be treated more carefully in the nano or micro-fluidic case as the  attractive region near the wall cannot be necessarily treated as simply being on the surface and there will typically be a thickness of the confined region which may  be comparable to the  width of the channel. As an example, if one considers electroosmotic flow, the flow is uniform at distances larger than the Debye screening length $\ell_D$ from the walls and decays to zero quickly near the walls. Even if the variation of the diffusion constant is neglected, if the tracer is charged, then it will interact with the walls leading to a considerable modification of its dispersion. 
{    \section{Summary of the main results}}
    Motivated by the observations made above, the goal of this paper is to present a general formula for the Taylor dispersion coefficient for channel-like systems which are uniform along the channel direction $x$ but inhomogeneous in the direction $z$ perpendicular to the channel. We will show that this formula leads to very simple analytic forms of the effective diffusivity in simple models describing the above mentioned situations. 
    
\begin{figure}[h!]
        \includegraphics[width=8 cm]{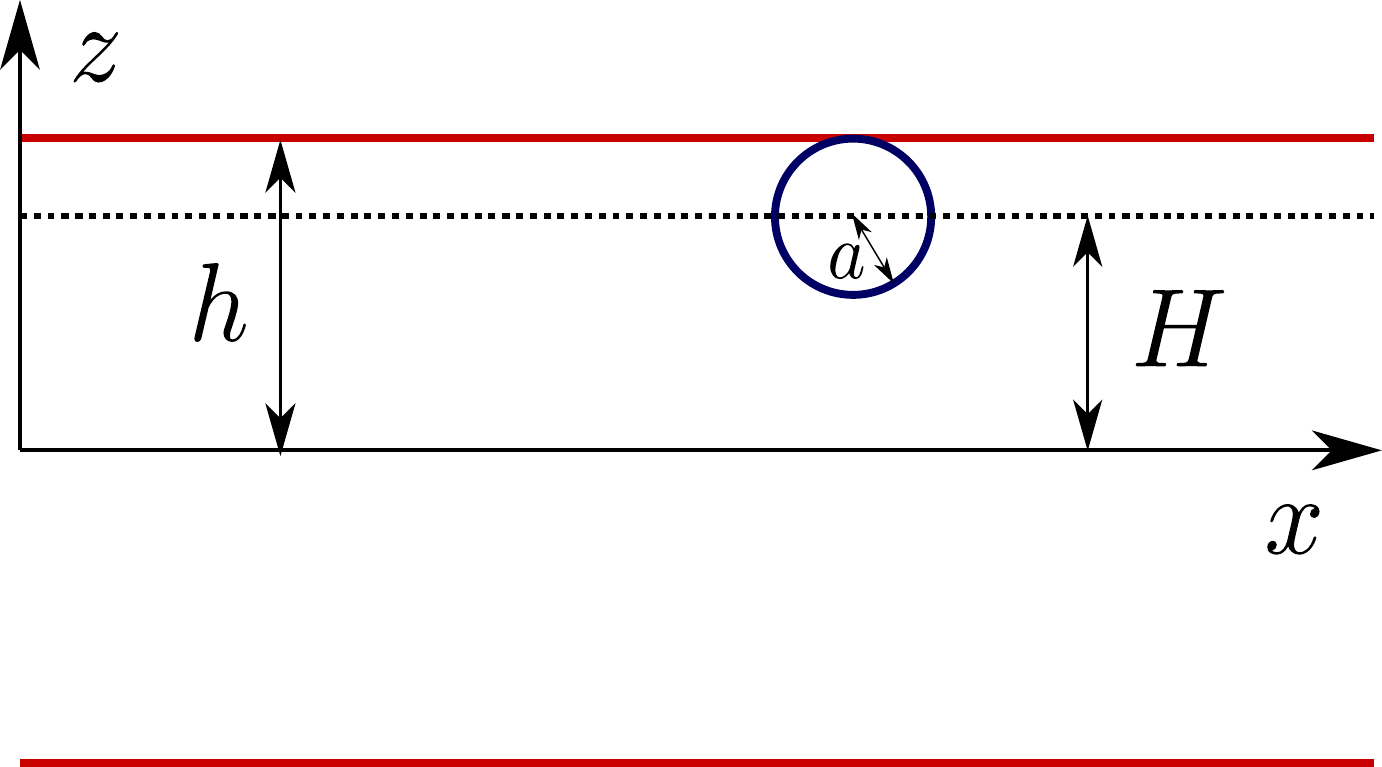}
        \caption{{ A schematic drawing representing the physical height $h$ and available height $H$. } }
        \label{dessinchan1}
    \end{figure}

    In the two dimensional channel problem, we assume that we have a channel with walls at {$z=\pm h$} with no-flux boundary conditions at the walls. {For a finite-sized tracer particle, we denote by $H$  the height available to the center of the particle, so for a hard spherical particle of radius $a$, the physical height $h$ of the channel and the available height are related by $ H=h-a$ as shown in Fig. (\ref{dessinchan1}). It is important to bear in mind that the boundary for the diffusion equation of the tracer particle is thus at $z=\pm H$ while the boundary concerning the fluid flow is at $z=\pm h$.}
    The local transport properties are  specified by
    \begin{eqnarray}
        D_\perp(z) &-& \ \text{tracer  diffusion constant perpendicular to the channel axis}\\
        D_{||}(z) &-& \  \text{tracer diffusion constant parallel to the channel axis}\\
        V(z) &-& \ \text{potential acting on the tracer}\\
        v(z) &-& \ \text{advection velocity  by hydrodynamic flow or force along the channel axis}
    \end{eqnarray} 
    The full advection-diffusion equation for the joint probability distribution $p(z,x;t)$ of the positions $Z_t$ and $X_t$ perpendicular to  and along the channel is therefore 
    \begin{align}
        \frac{\partial p(z,x;t)}{\partial t}
        =   & \frac{\partial }{\partial z}\left[D_\perp(z) \left(\frac{\partial p(z,x;t)}{\partial z} 
        +   \beta V'(z)p(z,x;t)\right)\right]\nonumber  \\
        &+ D_{||}(z)\frac{\partial^2 p(z,x;t)}{\partial x^2}-v(z)\frac{\partial p(z,x;t)}{\partial x},\label{fp1}
    \end{align}
    where $\beta = 1/(k_BT)$, the prime denotes the derivative with respect to $z$, $T$ is the temperature and $k_B$ is Boltzmann's constant.
    
    \noindent Integrating over $x$ gives the marginal probability distribution function $p(z;t)$ for $Z_t$ (we use the same notation $p$ but as it only has a single spatial argument, it is clear that the marginal distribution is intended): 
    \begin{equation}
        \frac{\partial p(z;t)}{\partial t} = \frac{\partial }{\partial z}\left[D_\perp(z) \left(\frac{\partial p(z;t)}{\partial z}
        +\beta V'(z)p(z;t)\right)\right] .\label{fp2}
    \end{equation}
    The steady state of this equation is given by the Gibbs-Boltzmann distribution:
    \begin{equation}
        p_0(z) = \frac{1}{\mathcal{N}} e^{-\beta V(z)},\label{EqDistr}
    \end{equation}
    where $\mathcal{N}$ is a normalization constant (the partition function):
    \begin{equation}
        \mathcal{N}=\int_{-H}^H dz \ e^{-\beta V(z)}. 
    \end{equation}
    The present paper deals with the late time transport along the channel, characterized by an effective drift $v_{e}$ and diffusion constant $D_{||e}$, given that the initial distribution of the tracer position in the $z$ direction is the equilibrium distribution (\ref{EqDistr}). The first moments of the random process $X_t$ read
    \begin{align}
        &\left< X_t\right>  =  v_{e}t \ &\text{for all}\ t>0 \\
        &\left< X^2_t\right>_c = \left< (X_t-\left< X_t\right>)^2 \right>  \simeq  2(D_{||e}t - C_{||e})  & \ \text{ as} \ t\to\infty,
    \end{align}
    where the initial position is $X_0=0$,  $\left< \cdot\cdot \right> $ denotes the ensemble average, and the notation $\left< \cdot\cdot \right>_c$ represents the connected part, so that $\left< X_t^2\right>_c$ is simply the variance of $X_t$. The result for the effective drift is valid for all times and is trivial to derive: 
    \begin{equation}
        v_e=\left< v\right>_0,
    \end{equation}
    where $\left<\cdot\cdot\right>_0$  denotes the average over $p_0$. 
    The effective diffusion coefficient admits a very simple analytical expression,
    \begin{equation}
        D_{||e}= \left< D_{||}\right>_0 + \frac{1}{\mathcal{N}}\int_{-H}^H dz \ \frac{J^2(z)e^{\beta V(z)}}{D_\perp(z)}\label{maind}.
    \end{equation}
    The first term is simply generated by molecular diffusion along the channel in the absence of any hydrodynamic flow, while the second is generated by Taylor dispersion due to the flow. The current term $J(z)$ depends only on the potential $V(z)$ and the flow field $v(z)$ and is simply given by
    \begin{equation}
        J(z) = \int_{-H}^z dz' e^{-\beta V(z')}[v(z')-\left< v\right>_0].\label{jtaylor}
    \end{equation}
    We thus see that the diffusion along the channel can be characterized by a simple integral expression for a wide variety of physical systems. The simple expression (\ref{maind}) can be derived by using Brenner's formalism for macrotransport processes \cite{fra89,bre93,bre76}, or Kubo-like formulas \cite{gue15a,gue15b}, here we will present a simple derivation in Section \ref{SectionDerivation}. We will also derive the corresponding general expression for Taylor dispersion in uniform cylindrical pipes, recovering simply a result given by Brenner and Gaydos \cite{bre76}. We will also give a general expression for the next order correction $C_{||e}$ to the mean squared displacement at large times,
    which turns out to be a constant term and which should be experimentally observable. We show below that
    \begin{equation}
        C_{||e} = \left<R^2\right>_0 - \left<R\right>_0^2 , \label{EqForCorrections}
    \end{equation}
    with
    \begin{equation}
        R(z)= \int_{-H}^z dz'\ \frac{J(z')e^{\beta V(z')}}{D_{\perp}(z')}.
    \end{equation}
    Note that Eq. (\ref{EqForCorrections}) indicates without ambiguity that $C_{||e} $ is positive,  this means that a linear fit of the average late time mean squared displacement $y(t) =\left< X_t^2\right>_c$ should cross the $y=0$ axis at positive value of $t$ and consequently presents a check of the quality of the statistical data acquired in experiments.
    
\begin{figure}[h!]
        \includegraphics[width=12 cm]{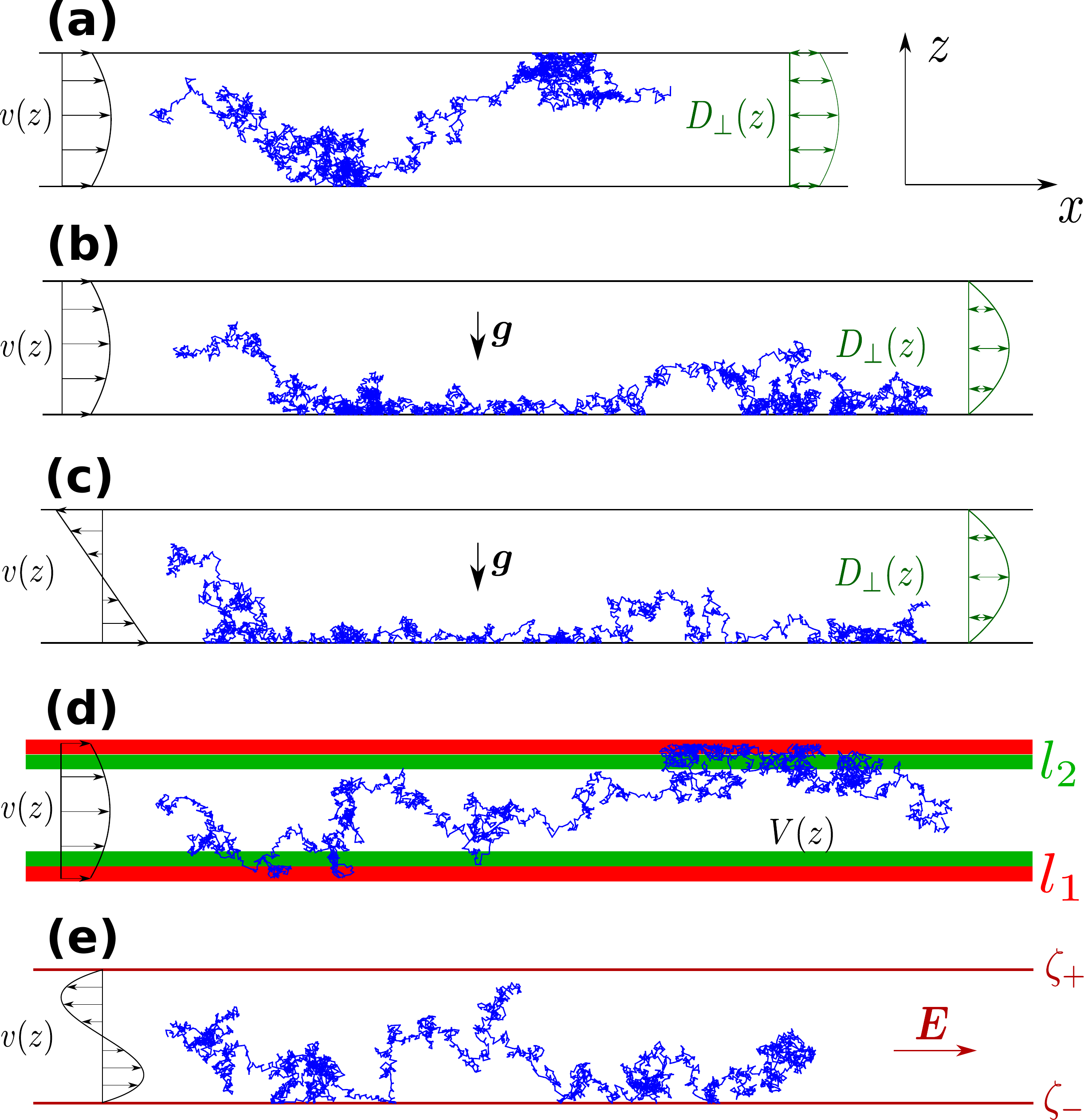}
        \caption{A schematic summary of the problems of Taylor dispersion addressed in this paper: (a) Systems with no wall interactions but with  diffusion constants in the height direction varying quadratically  (parabolic diffusion model) as a function of the height in the channel with a Poisseuille flow; (b) the same model as case (a), but restricted to zero height diffusion coefficient  at the surfaces, with an added gravitational potential; (c) the same model as (b) but with a Couette flow; (d) Poiseuille flow with strongly varying potentials near the surface, including attractive and repulsive components; (e) electroosmotic flow with different surface charges on each channel, the flow being the well known plug flow in the symmetric case. }
        \label{schema_channels}
    \end{figure}

    The main point of the paper is that the simple   formulas (\ref{maind},\ref{jtaylor})  can be applied to provide a number of simple, exact, analytical formulas in a number of applications. These applications are summarized schematically in Fig. \ref{schema_channels} and are described in the following text. In Section \ref{SectionParabolicDiff}, we will consider a simple model where the diffusion constant normal to the channel axis depends on the distance from the center in a quadratic fashion, a simple model of surface hindered diffusion considered before in the literature \cite{lau07,avn20}, though not in the context of Taylor dispersion. In this particular model, we also show how gravity acting normal to the channel modifies the dispersion due to the hydrodynamic flow. Then, in Section \ref{surf}, we will consider Taylor dispersion in a plane channel where there are interactions with the walls. We find a general expression when the interactions with the walls are strong and localised near the walls. It is then shown how the  formalism expounded here recovers and generalizes, in a straightforward and very physical way, a number of existing results for surface reaction diffusion models where the particle can be absorbed onto and desorb off the surface \cite{bre93,lev12,ber13}. 
    Finally, in Section \ref{SectionElectroOsmotic}, we apply the formalism to determine Taylor dispersion due to an electroosmotic flow, extending an existing result for a planar geometry \cite{gri99} to the case where the two confining surfaces have different surface charges or zeta potentials. 
    
    \section{Derivation of general results}\label{form}
    \label{SectionDerivation}
    \subsection{Dispersion in a two-dimensional uniform channel}
    We start by considering a two-dimensional channel, and we write the  Fokker-Planck equation Eq. (\ref{fp2}) for the tracer particle position $Z_t$ in the direction $z$ (perpendicular to the channel) as 
    \begin{equation}
        \frac{\partial p(z;t) }{\partial t} = - \mathcal{H} p(z;t),\label{fpz}
    \end{equation}
    and so 
    \begin{equation}
        \mathcal{H} = -\frac{\partial }{\partial z}\left(D_\perp(z)\left[ \frac{\partial }{\partial z}\cdot +\beta V'(z) \cdot\right]\right).
    \end{equation}
    No flux conditions \cite{gar09} at the upper and lower surfaces of the channel $z=\pm H$ lead to
    \begin{equation}
        \left[D_\perp(z) \left(\frac{\partial p(z;t)}{\partial z} +\beta V'(z) p(z;t) \right)\right]_{z=\pm H} =0. \label{noflux}
    \end{equation}
    
    \noindent To compute the dispersion in the $x$ direction, it is useful to represent the longitudinal process $X_t$, described by the Fokker-Planck equation (\ref{fp1}) in terms of the Langevin equation
    \begin{equation}
        \frac{dX_t}{dt} = \sqrt{2D_{||}(Z_t)}\eta(t) + v(Z_t),\label{sdex}
    \end{equation}
    where $\eta(t)$ is a Gaussian white noise of zero mean and with correlation function
    \begin{equation}
        \left< \eta(t)\eta(t')\right> = \delta(t-t').
    \end{equation}
    Note that there are no ambiguities in the prescription of the stochastic calculus to be used in Eq. (\ref{sdex}), as the noise is multiplicative in the position $Z_t$ but not in $X_t$. 
    
   \noindent Integrating and then averaging Eq. (\ref{sdex}) over the noise $\eta(t)$ and with respect to the initial position, $Z_{t=0}$ being assumed to be in equilibrium, yields the trivial result 
    \begin{equation}
        \left< X_t \right> = \left< v\right>_0t,
    \end{equation}
    and so the average drift has no time  dependence. Now,  squaring the integrated form of Eq. (\ref{sdex})  and carrying out the same averages yields 
    \begin{equation}
        \left< X^2_t \right>_c = 2\left< D_{||}\right>_0 t +\int_0^t ds \int_0^t ds' \left< [v(Z_s)-\left< v(Z_{s})\right> ][v(Z_{s'})-\left< v(Z_{s'})\right>] \right> \label{sdt}.
    \end{equation}
    We can decompose the mean squared displacement above as a purely diffusional term  (denoted with the { subscript} $D$ in { following expression})
    \begin{equation}
        \left< X^2_t\right>_{cD} = 2\left< D_{||}\right>_0 t,\label{do}
    \end{equation}
    and a term corresponding to the Taylor dispersion (denoted with the { subscript} $T$ { in the following expression})
    \begin{equation}
        \left< X^2_t\right>_{cT}=\int_0^t ds \int_0^t ds' \left< [v(Z_s)-\left< v(Z_{s})\right> ][v(Z_{s'})-\left< v(Z_{s'})\right>] \right>.
    \end{equation}
 { It is important to note here that if started from equilibrium in the height direction of the channel the only finite time correction to the longitudinal diffusion constant comes from the Taylor dispersion component, the contribution from pure diffusion is linear in time for all times.}
    
    Let us  denote by $p(z|z';t)$ the transition density to go from $z'$ at time zero to $z$ at time $t$ for the process $Z_t$, so $p(z|z';t)$ obeys Eq. (\ref{fpz}) with the initial condition $p(z|z';0)=\delta(z-z')$. With this notation, we find for the process $Z_t$ starting from equilibrium that
    {\begin{equation}
            \left< X_t^2\right>_{cT}= 2 \int_0^t ds \int_0^s ds'\int_{-H}^H dz\int_{-H}^H dz' v(z) v(z')\left[ p(z|z';s-s')p_0(z')-p_0(z)p_0(z')\right].\label{c42}
    \end{equation} }
    
    To proceed further, we consider the eigenvalues $\lambda$ of the operator $H$, associated to the left and right eigenfunctions $ \psi_{L\lambda}(z)$ and $ \psi_{R\lambda}(z)$ defined by 
    \begin{align}
        & \mathcal{H}\psi_{R\lambda}= \lambda \psi_{R\lambda},\\
        & \mathcal{H}^\dagger\psi_{L\lambda}= \lambda \psi_{L\lambda},
    \end{align}
    where $ \mathcal{H}^\dagger$ denotes the adjoint of $ \mathcal{H}$ (which in general is not self adjoint), and we impose that these eigenfunctions are normalized according to 
    \begin{align}
        \int_{-H}^H dz \  \psi_{L\lambda}(z)\psi_{R\lambda'}(z)=\delta_{\lambda,\lambda'}.
    \end{align}
    Then, the solution of  Eq. (\ref{fpz}) for $p(z|z';t)$ is given by
    \begin{equation}
        p(z|z';t)= \sum_\lambda \psi_{R\lambda}(z)\psi_{L\lambda}(z')e^{-\lambda t}.
    \end{equation}
    The right eigenfunctions obey the no flux boundary condition given in Eq. (\ref{noflux}) and one can show
    \cite{gar09} that the left eigenfunctions obey the Neumann boundary condition $\frac{d}{dz}\psi_{L\lambda}(z)|_{z=\pm H}=0$ at the boundaries. We note that, in principle, we could avoid left and right eigenvectors by making a transformation that renders $\mathcal{H}$ self adjoint, however it is actually easier to work with $\mathcal{H}$ as it stands as a forward Fokker-Planck operator. %(one can verify that doing this of course leads to the same result). 
    \noindent The eigenfunction corresponding to $\lambda=0$ can be written as $\psi_{R0}(z)= p_0(z)$ and $\psi_{L0}(z)= 1$
    and in this form they respect the normalization condition $\int dz \psi_{R0}(z)\psi_{L0}(z)= \int dz p_{0}(z)=1$.
    Using this representation of $p(z|z';t)$ in the Kubo formula Eq. (\ref{c42}), we find
    \begin{equation}
        \left< X_t^2\right>_{cT} = 2 \int_{-H}^H dz \int_{-H}^H dz' \sum_{\lambda>0}\left[\frac{t}{\lambda} -\frac{1}{\lambda^2} + \frac{\exp(-\lambda t)}{\lambda^2}\right]\psi_{R\lambda}(z)\psi_{L\lambda}(z')v(z)v(z') p_0(z').\label{c43}
    \end{equation}
    The method proposed above is known in the literature as the method of moments. When applied to standard Taylor dispersion, it can be formulated via the diffusion equation 
    \cite{bar83} or by direct computation of the moments\cite{ved14,gue15a,gue15b} as it is the case here. The late time behavior of the lowest moments can also be obtained using the powerful and extremely general macro-transport theory of Brenner and coworkers \cite{fra89,bre93} but also using invariant manifold theory \cite{mer90, mer94,bal95}.
    In principle, the method can be pushed further if the eigenfunctions and eigenvalues are known. However, even if they are known, one sees that the results are given in terms of infinite sums which may not always be obvious to compute. In general, therefore, the evaluation of the full temporal behavior of $\left< X_t^2\right>_{cT}$ requires the full solution of the  transition density $p(z|z';t)$ which is not usually available. However, we can extract the long time behavior in a generic fashion using a formulation in terms of Green's functions\cite{gue15a,gue15b} and which can be shown to be intimately linked to the macro-transport theory of Brenner \cite{bre93}.
    
    When $t\gg 1/\lambda_1$ where $\lambda_1$ is the first non-zero eigenvalue of $H$, Eq. (\ref{c43}) becomes
    \begin{equation}
        \left< X_t^2\right>_{cT} \simeq 2 D^{(T)}_{||e} t - 2 C^{(T)}_{||e},\label{latetime1}
    \end{equation}
    with 
    \begin{equation}
        D^{(T)}_{||e}=  \int_{-H}^H dz \int_{-H}^H dz' v(z)\sum_{\lambda>0}\frac{\psi_{R\lambda}(z)\psi_{L\lambda}(z')}{\lambda}v(z') p_0(z'),
    \end{equation}
    and 
    \begin{equation}
        C^{(T)}_{||e}=  \int_{-H}^H dz \int_{-H}^H dz' v(z)\sum_{\lambda>0}\frac{\psi_{R\lambda}(z)\psi_{L\lambda}(z')}{\lambda^2}v(z') p_0(z').
    \end{equation}
    Further progress can be made by noting that the quantity
    \begin{equation}
        G_P(z,z') = \sum_{\lambda>0}\frac{\psi_{R\lambda}(z)\psi_{L\lambda}(z')}{\lambda},
    \end{equation}
    is the pseudo-Green's function \cite{barton1989elements} of the operator $ \mathcal{H}$ , i.e. it satisfies 
    \begin{equation}
        \mathcal{H}G_P(z,z')=\delta(z-z') - p_0(z).
    \end{equation}
    This relation follows from the completeness relation $\sum\limits_{\lambda}\psi_{R\lambda}(z)\psi_{L\lambda}(z')=\delta(z-z')$. As a consequence, if we define the auxiliary field
    \begin{equation}
        f(z)= \int_{-H}^H dz' \ G_P(z,z') v(z')p_0(z'), \label{Def_f}
    \end{equation}
    we see that $f$ obeys 
    \begin{equation}
        \mathcal{H}f(z) = [v(z)-\left< v\right>_0]p_0(z).\label{eqf1}
    \end{equation}
    This then gives the compact formula
    \begin{equation}
        D^{(T)}_{||e}= \int_{-H}^H dz\  v(z) f(z) .\label{dle11}
    \end{equation}
    Now, solving for Eq. (\ref{eqf1}) with no flux boundary condition Eq. (\ref{noflux}) and the {\em orthogonality} condition $\int_{-H}^H dz\  f(z)=0$ (which is obvious in Eq. (\ref{Def_f})), we can find the explicit solution for $f$. A few algebraic manipulations (described in Appendix \ref{agen}), including integration by parts lead to the announced expression
    \begin{equation}
        D^{(T)}_{||e} = \frac{1}{\mathcal{N}}\int_{-H}^H dz \ \frac{J^2(z)e^{\beta V(z)}}{D_\perp(z)} , \label{dtaylor1}
    \end{equation}
    with 
    \begin{equation}
        J(z) = \int_{-H}^z dz' \exp(-\beta V(z'))[v(z')-\left< v\right>_0].
    \end{equation}
    
    Next, the constant $C^{(T)}_{||e}$, which characterizes the finite time corrections, can be obtained by noting that  it can also be expressed in terms of the pseudo-Green's function:
    \begin{equation}
        C^{(T)}_{||e}=  \int_{-H}^H dz\ \int_{-H}^H dz'\  v(z)G^2_P(z,z')v(z') p_0(z'),
    \end{equation}
    where we have used the operator notation
    \begin{equation}
        G^2_P(z,z') = \int_{-H}^H dz'' \ G_P(z,z'')G_P(z'',z').
    \end{equation}
    Using this property, a few algebraic manipulations (see Appendix \ref{agen}) lead to the simple expression for 
    \begin{equation}
        C^{(T)}_{||e} = \left<R^2\right>_0 - \left<R\right>_0^2 \label{eqc1},
    \end{equation}
    with
    \begin{equation}
        R(z)= \int_{-H}^z dz'\ \frac{J(z')e^{\beta V(z')}}{D_{\perp}(z')}.
    \end{equation}
    An alternative formula for $C^{(T)}_{||e}$ is given in Eq. (\ref{eqc}) of Appendix  \ref{agen}.
    Along with Eq. (\ref{do}), this leads to the main result Eq. (\ref{maind}). 
    From Eq. (\ref{dtaylor1}), we immediately see a number of well known characteristics of Taylor dispersion. First, we see that through its dependence on $J(z)$ is the variation of $v(z)$ which generates the dispersion, shifting $v(z)$ by a global constant velocity does not change $D_{||e}^{(T)}$. Interestingly, the current $J(z)$ is determined solely by the flow field and the potential.  We also see that Taylor dispersion   is amplified as $D_{\perp}$ is reduced, slowly diffusing particles are thus more strongly affected, a point that is well established.
    
    %{\bf The Fokker Plank equation in polar coordinates for the probability distribution $p(x,y,t)$ when written in terms of polar coordinates transforms via $p(x,y,t)dx dy= p(r,\theta,t)dr d\theta$ and so $p(x,y,t)=p(r,\theta)/r$. Inserting this into the Fokker-Planck equation for $p(x,y,t)$ for a systems with rotational symmetry we find
    %\begin{equation}
    %\frac{1}{r}\frac{\partial p(r,t)}{\partial t}= \frac{1}{r}\frac{\partial}{\partial r} D_{||}(r)r\left[\frac{\partial}{\partial r}
    %\frac{p(r,t)}{r} +\beta V'(r)\frac{p(r,t)}{r}\right],
    %\end{equation}
    %this then simplified to give the result below.}
    
    \subsection{Dispersion in a cylindrical pipe}
    In a cylindrical channel or pipe with rotational symmetry about the axis $r=0$, the diffusion equation for the marginal probability distrubtion $p(r,t)$ for the radial component of the diffusion is
    \begin{equation}
        \frac{\partial p(r;t)}{\partial t} = \frac{\partial}{\partial r}\left[ D_{\perp}(r)\left(\frac{\partial p(r;t)}{\partial r}  + \left[\beta V'(r)-\frac{1}{r}\right]p(r;t)\right)\right].\label{cyl}
    \end{equation}
    The problem of dispersion in the longitudinal direction was analysed by the method of macro-transport theory by Brenner and Gaydos \cite{bre76} and we show here how their result can be immediately recovered from our formalism.
    From Eq. (\ref{cyl}), we see that the effective potential for this cylindrical geometry is $V_c(r) = -k_BT\ln(r) + V(r)$, treating $r=0$ and $r=R$ ($R$ being the pipe radius) as having no-flux boundaries. We thus find, transcribing the  above results for a planar channel, that 
    \begin{equation}
        D_{||e} =\left< D_{||}\right>_0   +   \frac{1}{\mathcal{N}}\int_{0}^R dr \ \frac{J^2(r)e^{\beta V(r)}}{r D_\perp(r)} ,\label{dtaylorc}
    \end{equation}
    with 
    \begin{equation}
        J(r) = \int_{0}^r dr'\  r' e^{-\beta V(r')}[v(r')-\left< v\right>_0],
    \end{equation}
    where $\left< \cdot\right>_0 = \int_{0}^R dr r \cdot p_0(r)$. These results are in agreement with those of Ref.\cite{bre76}. In addition, we find that the next order late time correction  is given by
    \begin{equation}
        C^{(T)}_{||e} = \left<R^2\right>_0 - \left<R\right>_0^2 ,\label{eqc3D}
    \end{equation}
    with
    \begin{equation}
        R(r)= \int_{0}^r dr'\ \frac{J(r')e^{\beta V(r')}}{r' D_{\perp}(r')}.
    \end{equation}

    \noindent In the following sections, we use these results to derive analytical expressions of the effective diffusion constant and late time correction for different configurations (see Fig.  \ref{schema_channels}). 
    
    \section{Planar channels with parabolic diffusivity profiles}
    \label{SectionParabolicDiff}
    Here, we consider a very narrow channel containing a tracer which is comparable to the tracer size. We can assume that the diffusion constant varies as (see Fig. \ref{schema_channels}(a))
    \begin{equation}
        D_{\perp}(z)= D_{\perp 0}\left(1-\frac{z^2}{H_d^2}\right),\label{van1}
    \end{equation}
    and of course, physically, one must impose the restriction $H_d\geq H$ { as the diffusion constant should be non-zero as long as the sphere does not touch the wall}.
This general model takes into account  the possibility that the transverse diffusivity is not exactly zero at the boundary. The model with $H_d = H$ has been proposed theoretically \cite{lau07, avn20} {and essentially follows from symmetry and the fact that there must be a regime where the true diffusivity varies quadratically corresponding to particle sizes comparable to the channel width.   Experimental results also appear to show that in this regime, of comparable particle and channel sizes,  a quadratic dependence gives a fair fit \cite{fau94,duf01} of the behavior of $ D_{\perp}(z)$. The full exact behavior of the diffusion tensor between two parallel planes is still not fully solved, despite the fact that the diffusion tensor for a single plane was computed by Faxen \cite{hap91} early in the 1920s. The Faxen result for a single surface can actually be used to give an estimate for the diffusion constant  via a superposition approximation proposed by Oseen \cite{ose27}. However, such an approximation only works when the two surfaces are well separated and so their hydrodynamic interaction via the tracer can be neglected. We pursue this parabolic diffusion model as it can be pushed through analytically. However, we emphasize that  Eq. (\ref{dtaylor1}) is straightforward to evaluate numerically for arbitrary complicated analytical expressions for $D_\perp(z)$ or indeed numerical \cite{gan80} or experimental \cite{fau94,duf01,vil20} data for $D_\perp(z)$.
        
        \subsection{Poiseuille flow, parabolic diffusivity}
        
        First, we assume that  the longitudinal fluid flow is Poiseuille, possibly with a non-zero slip length, we write  
        \begin{equation}
            v(z)= v_0\left(1-\frac{z^2}{H_s^2}\right).
        \end{equation}
        In general, both $v_0$ and $D_{\perp 0}$ will depend on the effective channel {height} $H$ and the particle radius $a$, in particular as the Poiseuille flow will be modified near the particle. The quadratic approximation for the diffusion constant is certainly valid in the center of the channel, although close to the surface, numerical computations and approximations show that it varies more rapidly \cite{gan80,duf01}. However, if the size of this surface region is small with respect to the particle size and there is no attractive surface potential, this simple model will give us an idea of how Taylor dispersion is modified by the reduction in diffusion due to the tracer's finite size when it is close to the walls. 
        
        Using these forms for the local diffusion constant and flow field and taking $ V=0$, we obtain
        \begin{equation}
            D_{\parallel e}^{(T)} = \frac{v_0^2H^6 }{135 H_s^4 D_{\perp 0} } {\Delta}_{PP}\left(c=\frac{H_d}{H}\right),
        \end{equation}
        where $c$ is the dimensionless ratio of two lengths $c= H_d/ H$, and so physically we must have $c\geq 1$. The scaling function ${\Delta}_{PP}$ associated with this {\em Parabolic-diffusivity, Poiseuille-flow { (PP)} problem} is given by
        \begin{equation}
            {\Delta}_{PP}(c)= c^2\left[-8 +25 c^2-15c^4+15c(c^2-1)^2 \coth^{-1}(c) \right].
        \end{equation}
        Inspection of the function ${\Delta}_{PP}(c)$ shows that it
        decays monotonically on $[1,+\infty)$,  Taylor dispersion  increases when $c$ decreases. In the limit where the transverse diffusivity vanishes at the walls ($H_d=H$, so $c=1$), we obtain maximal dispersion
        \begin{equation}
            D_{\parallel e}^{(T)} = \frac{2v_0^2H^6 }{135 H_s^4 D_{\perp 0} }, \label{q1}
        \end{equation}
        while in the limit $c\rightarrow +\infty$, we find the standard Taylor dispersion result
        \begin{equation}
            D_{\parallel e}^{(T)} = \frac{8 v_0^2H^6 }{945 H_s^4 D_{\perp 0} },
        \end{equation}
        which corresponds to minimal dispersion in this model. { The reason for the enhancement of Taylor dispersion here is obvious, it is due to the slowing down of the overall diffusion perpendicular to the channel as can be seen from the general form of  Eq. (\ref{dtaylor1}).}
        
        In this model, the constant $C_{||e} $ can also be calculated by using our formalism [Eq. (\ref{EqForCorrections})], the result is 
        \begin{align}
            C_{||e} =\frac{H^8 v_0^2 }{1620 D_{\perp0}^2 H_s^4 }    {\Gamma}_{PP}(c),
        \end{align}
        with 
        \begin{align}
            {\Gamma}_{PP}(c)=& c^4 \Big(4 (36 - 65 c^2 + 30 c^4) - 
            15 \ c \ (-1 + c^2)^2 \Big\{ 4 \text{coth}^{-1}(c) [-1 + 3 c \text{coth}^{-1}(c) ]   \nonumber\\
            &+      3 \ln \frac{c-1}{c+1} \ln \frac{4c^2}{c^2-1} - 
            6    \left[ \text{Li}_2\left(\frac{-1+c}{2c}\right)  -\text{Li}_2\left(\frac{1+c}{2c}\right)\right]\Big\} \Big),
        \end{align}
        where $\text{Li}_2$ is the polylogarithm function. This formal expression can be simplified for a diffusivity vanishing at the walls $H_d=H$:
        \begin{align}
            C_{||e}=\frac{H^8 v_0^2}{405 D_{\perp0}^2 H_s^4},
        \end{align}
        whereas for the standard case of uniform lateral diffusivity, one finds 
        \begin{align}
            C_{||e}=\frac{4 H^8 v_0^2}{4725 D_{\perp0}^2 H_s^4}.
        \end{align}
 { We thus see that the finite time correction, as measured by the magnitude of $C$ is more important in the case where the diffusion constant varies in a parabolic fashion}.       
        
        %If we carry out the computation for a general form of $D_{\perp}(z)$ but keep the  Poiseuille form for the flow field we find
        %\begin{equation}
        %D_{||e}^{(T)}=\frac{v_0^2H^3}{18 H^4_s}\int_{-H}^H dz \frac{z^2(1-\frac{z^2}{H^2})^2}{D_\perp(z)}.
        %\end{equation}
        
        \subsection{Poiseuille flow, parabolic diffusivity with gravity}
        
        We now consider the case where the tracer particle is subject to a gravitational field  (see Fig. \ref{schema_channels}(b)) 
        and there is a density difference $\Delta \rho$ with the fluid. In order to avoid too many length scales, we consider a flow which can slip at the surface with $H_s > h$ but we assume that $H_d=H$ so
        the diffusion constant perpendicular to the channel vanishes { when the tracer is in contact with the wall}. In this case,  we can write
        \begin{equation}
            V(z) = g\ \Delta \rho\  \Omega \ z = k_BT z/\ell_B,
        \end{equation}
        where $\Omega$ is the volume of the spherical particle and we have used $\beta V(z) = z/\ell_B$ where $\ell_B$ is the Boltzmann length. We also  introduce the  dimensionless variable $c$ such that $H=c \ell_B$, so a large value of $c$ corresponds to the regime where the gravitational has a strong effect on the tracer position. The late time behavior of the Taylor dispersion component of the diffusion constant along the channel is given by
        \begin{equation}
            D_{||e}^{(T)}= \frac{ v_0^2 H^6}{D_{\perp 0}H_s^4} {\Delta}_{PPG}\left(c=\frac{H}{\ell_B} \right),
        \end{equation}
        where 
        \begin{equation}
            {\Delta}_{PPG}(c)= \frac{5 + 4 c^2 - 5 \cosh(2 c) + c\sinh(2c) + 4c\coth(c)\int_0^{2c}dt\ \frac{\cosh(t)-1}{t}}{c^4\sinh^2(c)} \label{delta_ppg} ,
        \end{equation}
        is the scaling function for {\em Parabolic-diffusivity, Poiseuille-flow with Gravity} { (PPG)}.
        The results Eq. (\ref{q1}) (for $H_d=H$)  in the absence of gravity are recovered from the expansion for small 
        $c$ which gives
        \begin{equation}
            D_{||e}^{(T)} \simeq \frac{ 2v_0^2 H^6}{135  D_{\perp 0}H_s^4}\left[1 + \frac{5}{14} c^2 - \frac{71}{525} c^4 + \frac{29}{990} c^6 +{\cal O}(c^8)\right].
        \end{equation}
        Interestingly, the function ${\Delta}_{PPG}(c)$ is non-monotonic, it increases as $c$ is increased before  achieving a single maximum at $c_{max}\simeq 1.69$ and then decays to  zero as 
        \begin{equation}
            {\Delta}_{PPG}(c)\simeq \frac{2}{c^3},
        \end{equation}
        the ultimate reduction in dispersion due to the particle being stuck near the lower wall when $c$ is large is easy to understand, whereas the enhancement of dispersion for small $c$ is more surprising. { The quadratic dependence for small $c$ comes from the symmetry upon reversing the direction of the gravitational or buoyancy force. }The function ${\Delta}_{PPG}$ is plotted in Fig. \ref{fig:ppsubfigure1}.
        \begin{figure}[h!]
        \centering
        \subfigure[Function $\Delta_{PPG}(c)$ - Eq. (\ref{delta_ppg})]{\includegraphics[width=8cm]{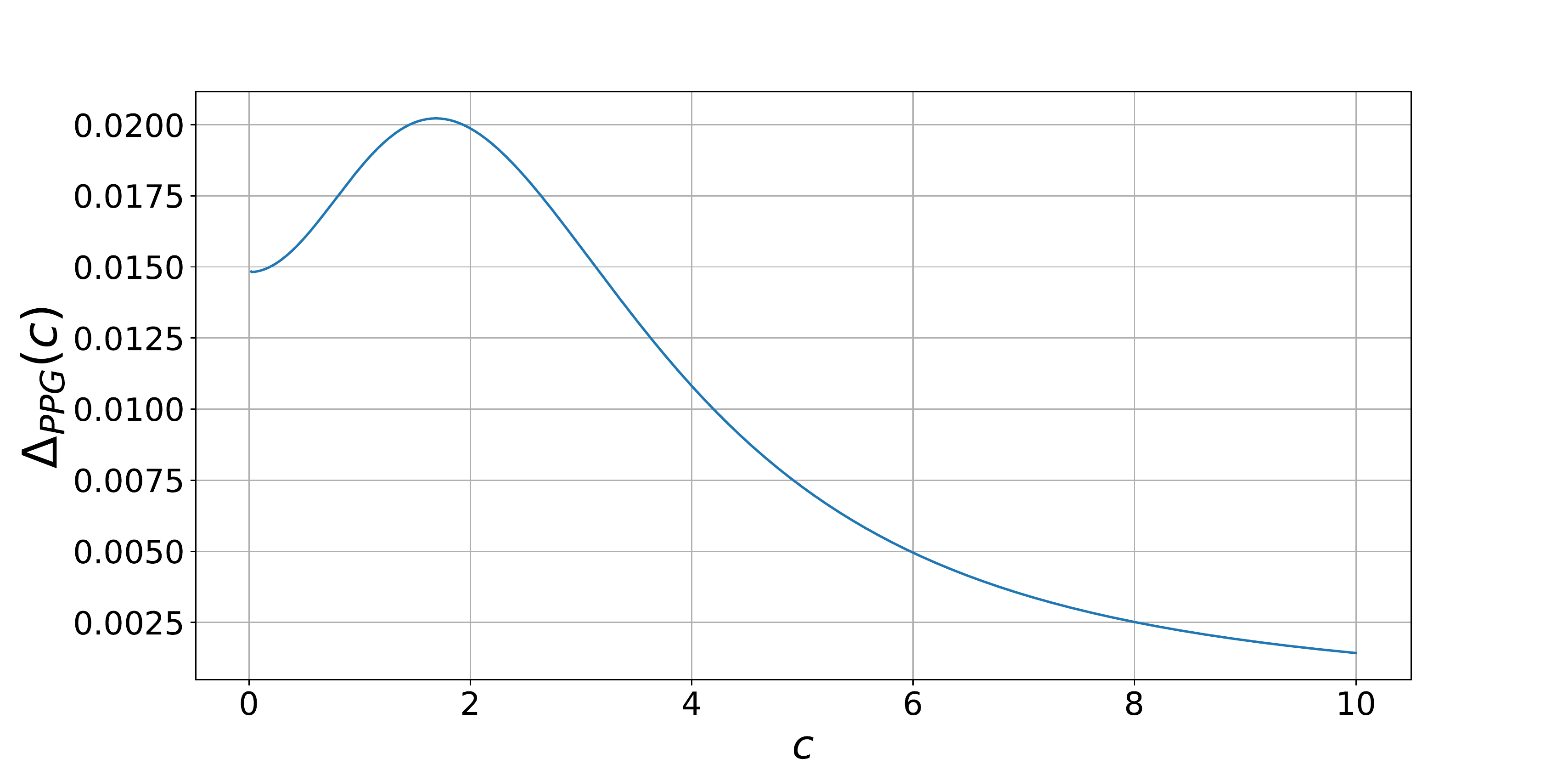}
            \label{fig:ppsubfigure1}}
        \quad
        \subfigure[Function $\Delta_{PCG}(c)$ - Eq. (\ref{delta_pcg})]{ \includegraphics[width=8cm]{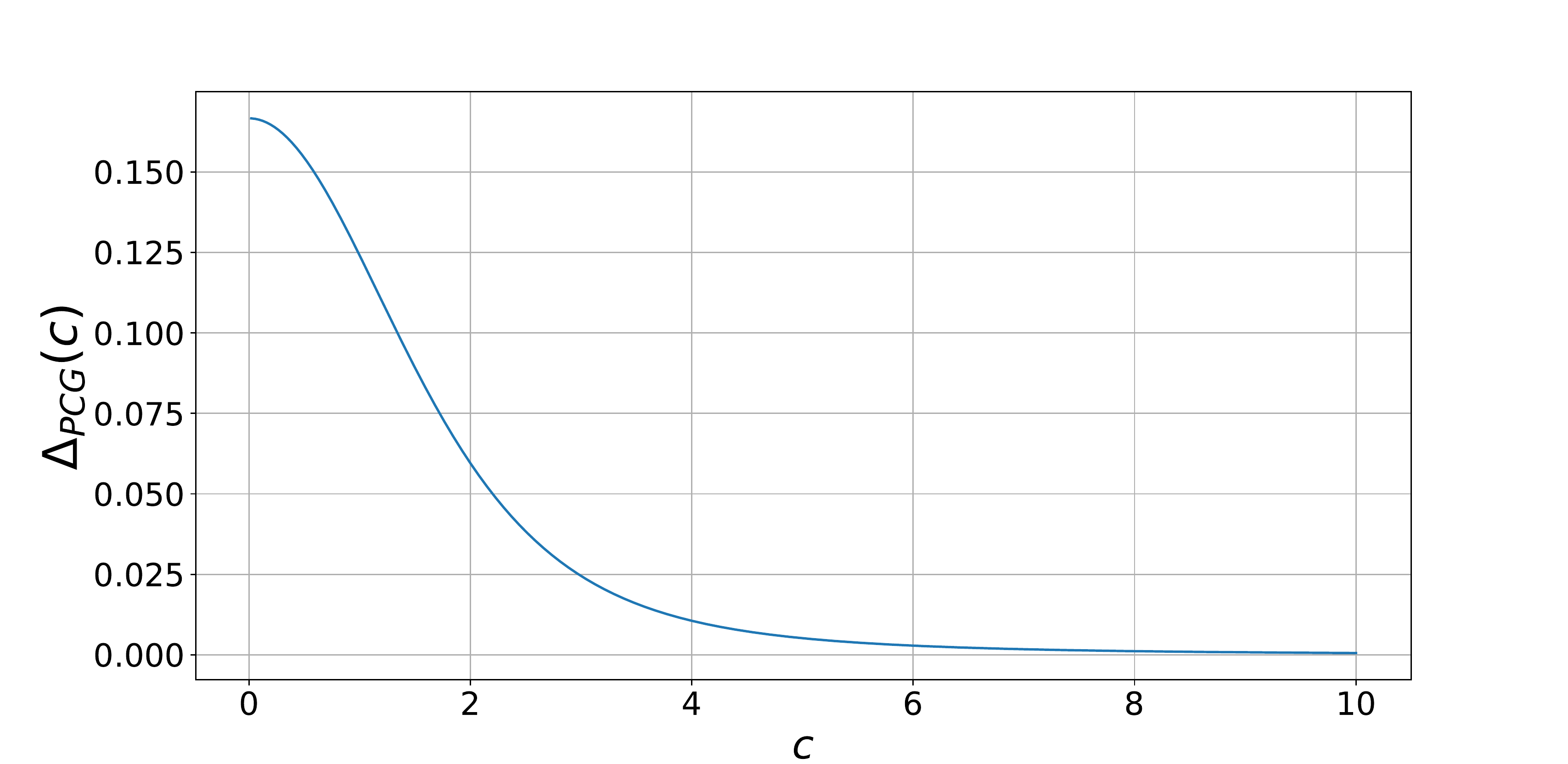} \label{fig:ppsubfigure2}}
        \caption{{The scaling functions ${\Delta}_{PPG}(c)$ (a) and ${\Delta}_{PCG}(c)$ (b) describing the effect of gravity on the Taylor dispersion component of the diffusion constant parallel to the channel for a tracer with a parabolically diffusivity perpendicular to the channel (in the case where it vanishes at the wall) and subject to gravity (increasing as $c$ increases). In Fig. (a), a Poiseuille flow is considered. Interestingly, the effect of turning on gravity initially increases the dispersion { while for a Couette flow (b), dispersion decreases monotonically. Dispersion is ultimately suppressed for large gravitational forces in both cases}.}}
        \label{fig:figure}
        \label{figpp}
    \end{figure}
        
        We can also compute the finite time corrections for this model, for which we give only the first terms of the small gravity expansion:
        \begin{equation}
            C_{||e}^{(T)} \simeq  \frac{H^8 v_0^2}{405\ D_{\perp0}^2 H_s^4}\left(1+  \frac{103}{84} c^2-\frac{ 17443 }{31500}  c^4+ {\cal O}(c^6)\right) ,
        \end{equation}
                
        \subsection{Couette flow and parabolic diffusivity, with/without gravity}
        
        If we consider a Couette flow with $v(z)=v_0 z/H_c$ in the absence of any potential and for an arbitrary transverse diffusion profile $D_\perp(z)$, we find
        \begin{equation}
            D_{||e}^{(T)}=\frac{v_0^2H^3}{8 H^2_c}\int_{-H}^H \frac{dz}{D_\perp(z)}\left(1-\frac{z^2}{H^2}\right)^2,
        \end{equation}
        and so for the quadratic diffusion profile given in Eq. (\ref{van1}), where we have again set $H_d=H$,  we find the simple result
        \begin{equation}
            D_{||e}^{(T)}=\frac{v_0^2H^4}{6 H_c^2 D_{\perp 0}}.
        \end{equation}
        The corresponding expression for the finite time correction reads
        \begin{equation}
            C_{||e}^{(T)}=\frac{H^6 v_0^2}{12 D_{\perp0}^2 H_c^2}.
        \end{equation}
        
         In the presence of a gravitational field (see Fig. \ref{schema_channels}(c)), we find
        \begin{equation}
            D_{||e}^{(T)}= \frac{ v_0^2 H^4}{H_c^2D_{\perp 0}} {\Delta}_{PCG}(c),
        \end{equation}
        where
        \begin{equation}
            {\Delta}_{PCG}(c) = \frac{c\cosh(c) \int_0^{2c} dt \frac{\cosh(t)-1}{t} - \sinh^3(c)}{c^2 \sinh^3(c)} \label{delta_pcg},
        \end{equation}
        is the scaling function for {\em Parabolic-diffusivity, Couette flow with Gravity} { (PCG)}.
        For small $c$ we find that 
        \begin{equation}
            {\Delta}_{PCG}(c)\simeq \frac{1}{6}-\frac{7 c^2}{135}+\frac{41 c^4}{3780}-\frac{19 c^6}{10125}+{\cal }O(c^8),
        \end{equation}
        while for large $c$
        \begin{equation}
            {\Delta}_{PCG}(c)\simeq \frac{1}{c^2}.
        \end{equation}
        In contrast  to what is seen for Poiseuille flow, for Couette flow, the function  ${\Delta}_{PCG}(c)$ is monotonically decreasing {(see Fig. \ref{fig:ppsubfigure2})} and so the effect of gravity slows down Taylor dispersion for all applied gravitational fields. 
    
        \section{Revisiting Taylor dispersion with strong wall interactions - relation with surface reaction models}
        \label{surf}
        
        The presence of attractive surface interactions can lead to tracer particles becoming localized at the surfaces. Physically, this localization is due to the presence of a potential. However, when the interactions are highly localized about the surface, one can use reaction diffusion models to model the dynamics near the surface \cite{bre93,lev12,ber13}. We show how the formalism developed here can be used to recover results of such reaction diffusion models.
        
        Let  us consider the  case where a square { well} potential is localized in a region near the walls, near which the particle can be reversibly bound (see Fig. \ref{schema_channels}(d)). {  Specifically, the potential is $-U$ at a distance less that $l_1$ from the wall. It is also possible that there is an energy barrier against being absorbed onto the wall, to model this we include  a region adjacent to the well of width  $l_2$ with a potential barrier of height $V$ (possibly vanishing)}. We will consider the symmetric case, where the upper and lower boundaries are identical (see Fig. \ref{figbarrier}(a)), and the non-symmetric one (see Fig \ref{figbarrier}(b)).

        In all these situations, we consider that the transverse diffusion in the bulk is constant and reads $D_b$. On the surface (for $-H<z<-H+l_1+l_2$ and $H-l_1-l_2<z<H$), the diffusion coefficients parallel and perpendicular to the channel axis read respectively $D_{s \parallel}$ and $D_{s \perp}$ {and they can be written in terms of the bulk diffusion constant $D_b$ via }$s_{\parallel} = D_{s \parallel} / D_b$ and $s_{\perp} = D_{s \perp} / D_b$).
        \begin{figure}[h!]
            \includegraphics[width=12cm]{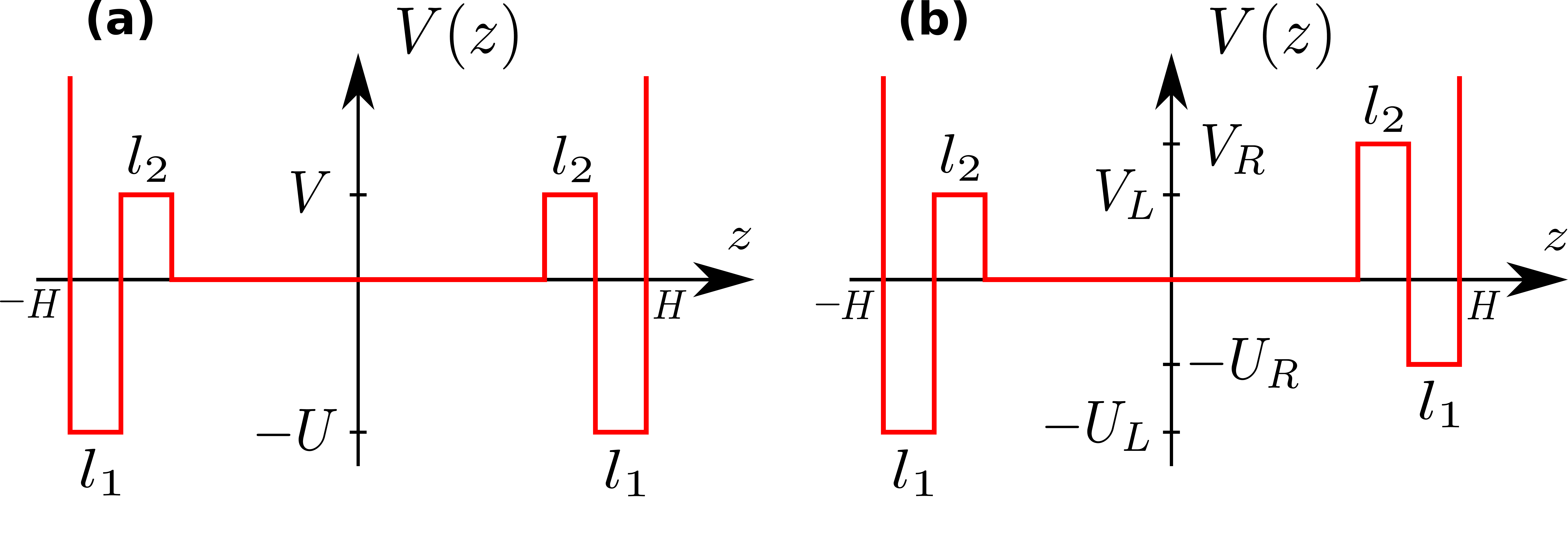}
            \caption{Surface potential for the two configurations discussed in the text, (a) is the symmetric configuration where both walls have identical potentials and  (b) shows the non-symmetric case where  the walls have differing potentials }\label{figbarrier}
        \end{figure}
        
        To start with, we consider the first configuration, represented in  Fig. \ref{figbarrier}(a). 
        %The Taylor dispersion contribution reads:
        %\begin{equation}
        %D^{(T)}_{||e} = \frac{1}{\cal N} \left(\frac{e^{\beta U_0}}{D_{s \perp}}\int_{-H}^{-H+l_1} J_1^2(z) dz + \frac{1}{D_{b}}\int_{-H+l_1}^{H-l_1} J_2^2(z) dz +\frac{e^{\beta U_0}}{D_{s \perp}}\int_{H-l_1}^{H} J_3^2(z) dz \right) \label{deff_int}
        %\end{equation}
        %where $J_1$, $J_2$ and $J_3$ are the current functions in the regions $ -H < z< -H +l_1$, $ -H+l_1 < z< H -l_1$, $ H-l_1 < z< H $ respectively. 
        We use the following expression of the flow
        \begin{equation}
            v(z) = v_0 \left(1- \frac{z^2}{H_s^2} \right)
        \end{equation} (with $H_s = {h}$ corresponds to a no-slip boundary condition and $H_s > {h}$ corresponds to a finite slip-length) to compute the long-time dispersion. By taking the limits: $a = l_1/H \rightarrow 0$ and $b = l_2/H \rightarrow 0$ while keeping the terms $ \alpha = a e^{\beta U}$ and $ \gamma = be^{\beta V}$ constant, we find the expression of the effective diffusivity:
        \begin{equation}
            \frac{D_{\parallel e}}{D_b} = \frac{1+\alpha s_{\parallel}}{1+ \alpha}+ \frac{ Pe^2}{(1+ \alpha)^3} \left(\frac{2}{105}+ \frac{6}{35}\alpha + \frac{17}{35}\alpha^2 +\frac{\alpha^2 \gamma}{s_{\perp}} \right) \label{sgen1}
        \end{equation}
        %\begin{equation}
        %\frac{D_{\parallel e}}{D_b} = \frac{1+\alpha s_{\parallel}}{1+ \alpha}+ \frac{ Pe^2}{(1+ \alpha)^3} \left(\frac{2}{105}+ \frac{6}{35}\alpha + \frac{17}{35}\alpha^2 \right)
        %\end{equation}
        with the Peclet number
        \begin{equation}
            Pe = \frac{2 v_0 H^3}{3H_s^2 D_b}.
        \end{equation}
        { We  notice that the potential barrier between the surface and the bulk can be removed by setting $\gamma=0$ and that this term always increases the Taylor dispersion.}
        %On can verify that  we recover the standard Taylor dispersion by taking $\alpha \rightarrow 0$. We now generalize  our model by considering the second, again symmetric, configuration (see Fig. \ref{figbarrier}(b)). After computing the integrals, taking the limits: $a = l_1/H \rightarrow 0$ and $b = l_2/H \rightarrow 0$ and keeping the terms $ \alpha = a e^{\beta U_1}$ and \textcolor{blue}{$ \gamma = be^{\beta V_1}$} constants, we find:
        %\begin{equation}
        %\frac{D_{\parallel e}}{D_b} = \frac{1+\alpha s_{\parallel}}{1+ \alpha}+ \frac{ Pe^2}{(1+ \alpha)^3} \left(\frac{2}{105}+ \frac{6}{35}\alpha + \frac{17}{35}\alpha^2 +\textcolor{blue}{\frac{\alpha^2 \gamma}{s_{\perp}}}  \right). \label{sgen1}
        %\end{equation}
        %The above show that the presence of a maximum before the wall leads to an enhancement of Taylor dispersion corresponding to the appearance of term proportional to $\gamma$.
        The above Eq. (\ref{sgen1}) can be explicitly compared to various models based on a reaction diffusion process between the surface and the bulk.
        The most general of these models can be described by a coupled set of reaction diffusion equations between the bulk density $p_b(z,x,t)$ and the surface density $p_s(x,t)$ (at the surface). For example,  for 
        channels symmetric about $z=0$  and with isotropic and  constant bulk and surface diffusion coefficients \cite{lev12}, one has 
        \begin{eqnarray}
            \frac{\partial p_b(z,x;t)}{\partial t} &=& D_b \nabla^2 p_b(z,x;t)  -v(z)\frac{\partial p_b(z,x;t) }{\partial x} \label{reac1} \\
            \frac{\partial p_s(x;t)}{\partial t} &=& D_s \frac{\partial^2 p_s(x;t)}{\partial x^2}-v_s \frac{\partial p_s(x;t)}{\partial x} -k_d p_s(x;t) + k_a p_b(H,x;t).\label{reac2}
        \end{eqnarray} 
        The term $k_d$ represents the rate at which the particle detaches from the wall and $k_a$ represents the
        rate of absorption at the wall from the neighboring bulk layer. The terms $D_s$ and $v_s$ represent the surface diffusion constant and surface flow or drift respectively. The boundary conditions for the bulk equation are easily derived from probability conservation and give 
        \begin{equation}
            D_b \frac{\partial p_b(H,x;t)}{\partial z}-k_d p_s(x;t) + k_a p_b(H,x;t)=0.\label{flux}
        \end{equation}
        Variants of the above system of Eqs. (\ref{reac1}-\ref{flux}) have been studied by various authors \cite{bre93,lev12,ber13}. In such  problems, the equilibrium distributions $p_{b}^{eq}(x) = p_{b}^{eq}$ and $p_{s}^{eq}(x)= p_{s}^{eq}$ are constant and obey Henry's law \cite{bre93}
        \begin{equation}
            p_{seq} =\frac{k_a}{k_d} p_{beq}, \label{henry}
        \end{equation}
        although this cannot be used as a boundary condition in the time dependent problem unless the flux  term proportional to $D_b$ in Eq. (\ref{flux}) is ignored (corresponding to very rapid reaction terms with respect to diffusion). The discussion of the boundary conditions and concepts such as partial reflection and absorption of Brownian motion is somewhat involved to analyse mathematically \cite{singer2008partially} and indeed to simulate numerically.  The advantage of the method used here is that the Brownian motion is standard and  the dispersion  can be derived on purely physical grounds based on local diffusivities and potentials.
        It was shown \cite{lev12} that for the model based on  Eqs. (\ref{reac1}-\ref{flux}), the result is given exactly by Eq. (\ref{sgen1}) with $\alpha = \frac{k_a}{k_d H}$ and $\frac{\gamma}{s_{\perp}}=\frac{D_b}{k_a H}$. Therefore, we can determine the expressions for the adsorption and desorption rates for the potential under consideration
        \begin{align}
            &k_a =\frac{D_{s\perp}}{l_2} e^{-\beta V}  \\
            &k_d =\frac{D_{s\perp}}{l_1 l_2} e^{-\beta (U+V)}.
        \end{align}
        We thus find that the absorption and desorption rate follow  Arrh\'enius laws, since $V$ and $V+U$ are the barriers that respectively limit the binding and the unbinding kinetics.  
        
        By taking the limit where $D_{s \perp}\to +\infty$}, the term proportional to $\gamma$ in Eq. (\ref{sgen1}) vanishes and the result becomes equivalent to that given by Brenner and Edwards \cite{bre93}. A similar calculation can be carried out for cylindrical pipes and allows one to recover the results for surface-mediated diffusion \cite{lev12,ber13}, where in addition the effect of a distinct surface flow (so not necessarily continuous with respect to the bulk flow) was also incorporated. Note that we can can also derive Eq. (\ref{sgen1}) by considering a more general, continuous and differentiable, potential and using saddle-point approximation (see Appendix \ref{AppendixWall}). 
    The finite time correction in the case can also be computed and reads
    \begin{equation}
        C_{e\parallel} = \frac{Pe^2 H^2}{(1+\alpha)^4} \left(\frac{1}{525}+\frac{4}{175}\alpha +\frac{8}{75} \alpha^2 + \frac{62}{315} \alpha^3+\frac{2}{15}\frac{\alpha^2\gamma}{s_{\perp}}+\frac{4}{5}\frac{\alpha^3\gamma}{s_{\perp}} +\frac{\alpha^3\gamma^2}{s_{\perp}^2} \right).
    \end{equation}
    
    We can also extend the previous result (\ref{sgen1}) to the case of an asymmetric surface potential (see Fig. \ref{figbarrier}.(b)). By taking the limits: 
    $ a=l_1/H \rightarrow 0 , \  b=l_2/H \rightarrow 0$ and keeping $ \alpha_i = a e^{\beta U_i}$ and $\gamma_i =b e^{\beta V_i} $ constants ($i \in \{L,R \}$), we find
    \begin{equation}
        \begin{split}
            \frac{D_{\parallel e}}{D_b}& = \frac{ 1+  \frac{s_{\parallel}\left(\alpha_L + \alpha_R \right)}{2}}{1+\frac{\alpha_L + \alpha_R}{2} } + \frac{Pe^2}{\left(1+ \frac{\alpha_L + \alpha_R}{2} \right)^3}\\ &\times \left(\frac{2}{105} + \frac{3}{35}\left(\alpha_L + \alpha_R \right) + \frac{13}{35} \left(\alpha_L^2 + \alpha_R^2 \right)- \frac{9}{35}\alpha_L\alpha_R  + \frac{1}{2 s_{\perp}}\left(\alpha_L^2\gamma_L +\alpha_R^2\gamma_R \right) \right).
        \end{split} \label{asym}
    \end{equation}
    
   \noindent One thus sees that when  $U_L=U_R$ and $V_L=V_R$, we recover Eq. (\ref{sgen1}). It is interesting to notice the appearance of a term  $\alpha_L \alpha_R$ representing a correlation effect between the two surfaces at $z= \pm H$.
    \begin{figure}[h!]
            \includegraphics[width=8cm]{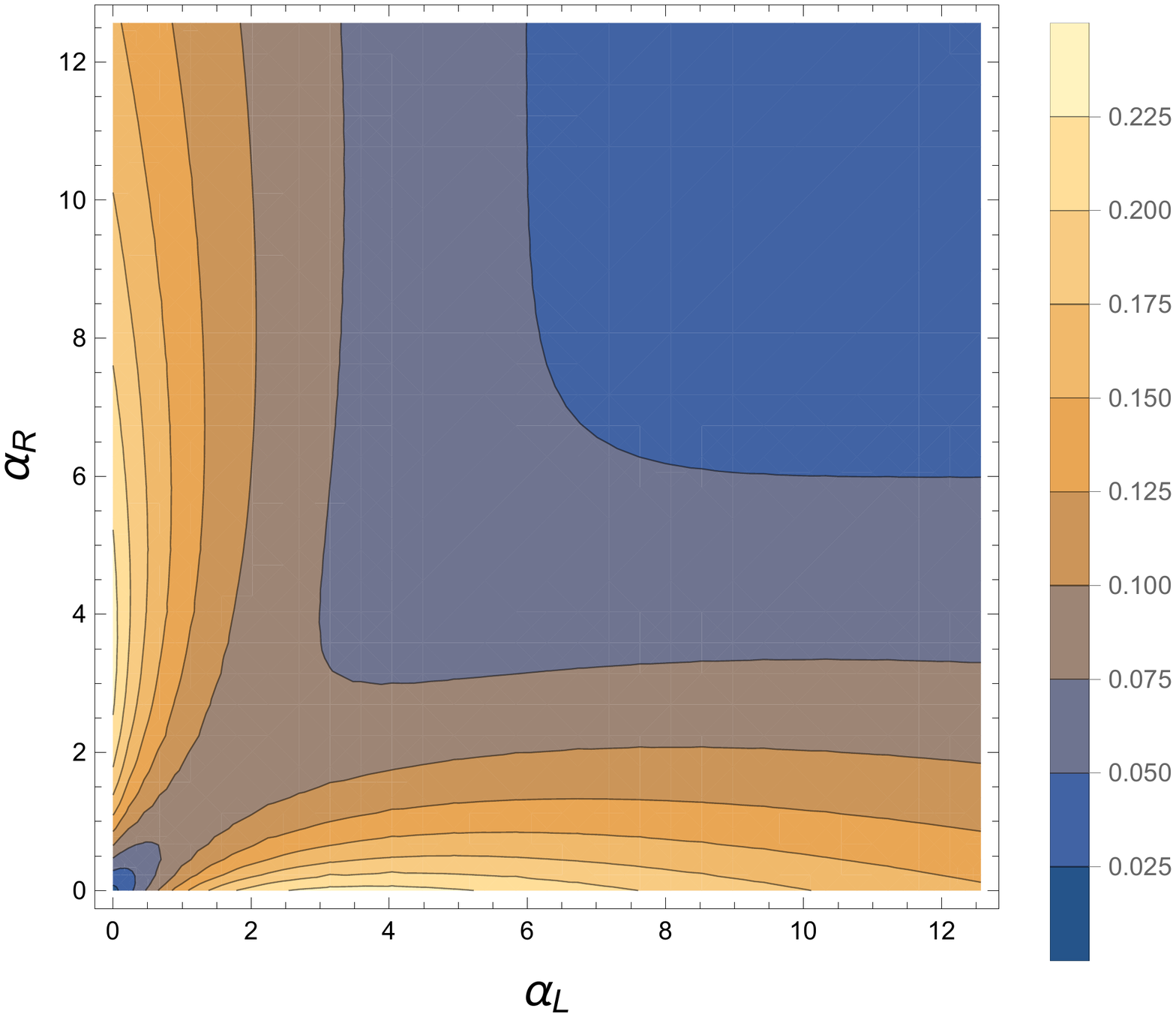}
            \caption{{The  function $\Delta_{SP}(\alpha_R,\alpha_L)$ given by Eq. (\ref{dsp}) describing the Taylor dispersion due to two, in general different absorbing walls, but with no barrier before the walls 
            (so $\gamma_R=\gamma_L=0$). For smaller values of $\alpha_R$ and $\alpha_L$ the absorption enhances Taylor dispersion, particularly in the regions where one is small with respect to the other.}}\label{spot}
        \end{figure}
    { In the case where $\gamma_L=\gamma_R=0$ the contribution to the diffusion constant from Taylor dispersion can be written as 
    \begin{equation}
    \frac{D^{(T)}_{\parallel e}}{D_b}= Pe^2 \Delta_{SP}(\alpha_R,\alpha_L), 
    \end{equation}
    with 
\begin{equation}
\Delta_{SP}(\alpha_R,\alpha_L)= \frac{ \frac{2}{105} + \frac{3}{35}\left(\alpha_L + \alpha_R \right) + \frac{17}{140} \left(\alpha_L + \alpha_R \right)^2 +\frac{1}{4}(\alpha_L-\alpha_r)^2  }{(1+ \frac{\alpha_L + \alpha_R}{2})^3}.\label{dsp}
\end{equation}
When $\alpha_1$ and $\alpha_2$ become large the dispersion is reduced due to the term in the denominator. However,  for smaller values, and especially differing values of $\alpha_R$ and $\alpha_L$, there is clearly an enhancement of Taylor dispersion due to absorption at the walls as seen in Fig. \ref{spot} where the function $\Delta_{SP}(\alpha_R,\alpha_L)$ is plotted.

}
    \section{Taylor dispersion in an electroosmotic flow}
    \label{SectionElectroOsmotic}
\begin{figure}[h!]
        \centering
        \subfigure[Symmetric function $\Delta_s(c)$ - Eq. (\ref{fsf})]{\includegraphics[width=7.5cm]{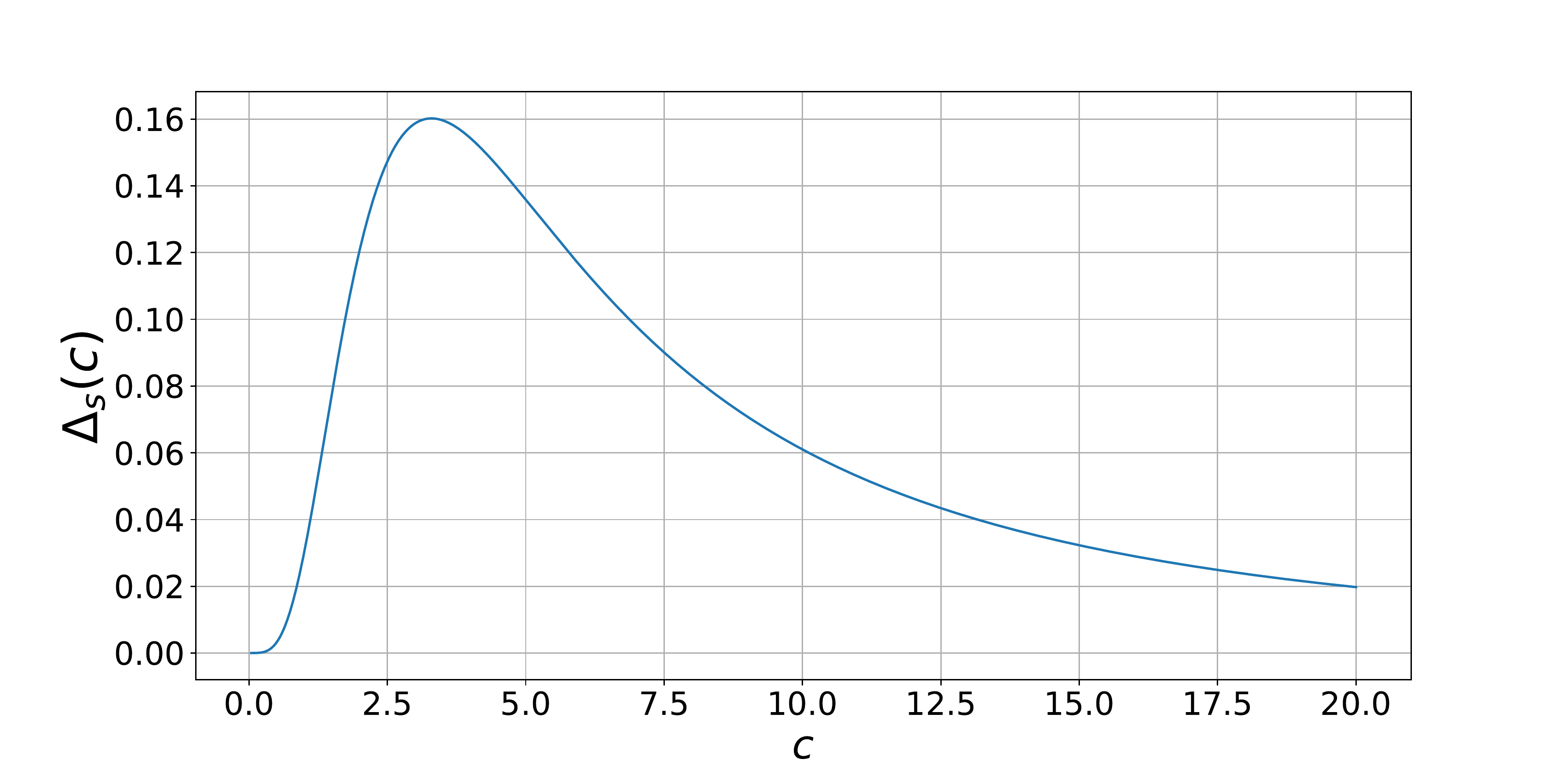}
            \label{fig:subfigure1}}
        \quad
        \subfigure[Antisymmetric function $\Delta_a(c)$ - Eq. (\ref{faf})]{ \includegraphics[width=7.5cm]{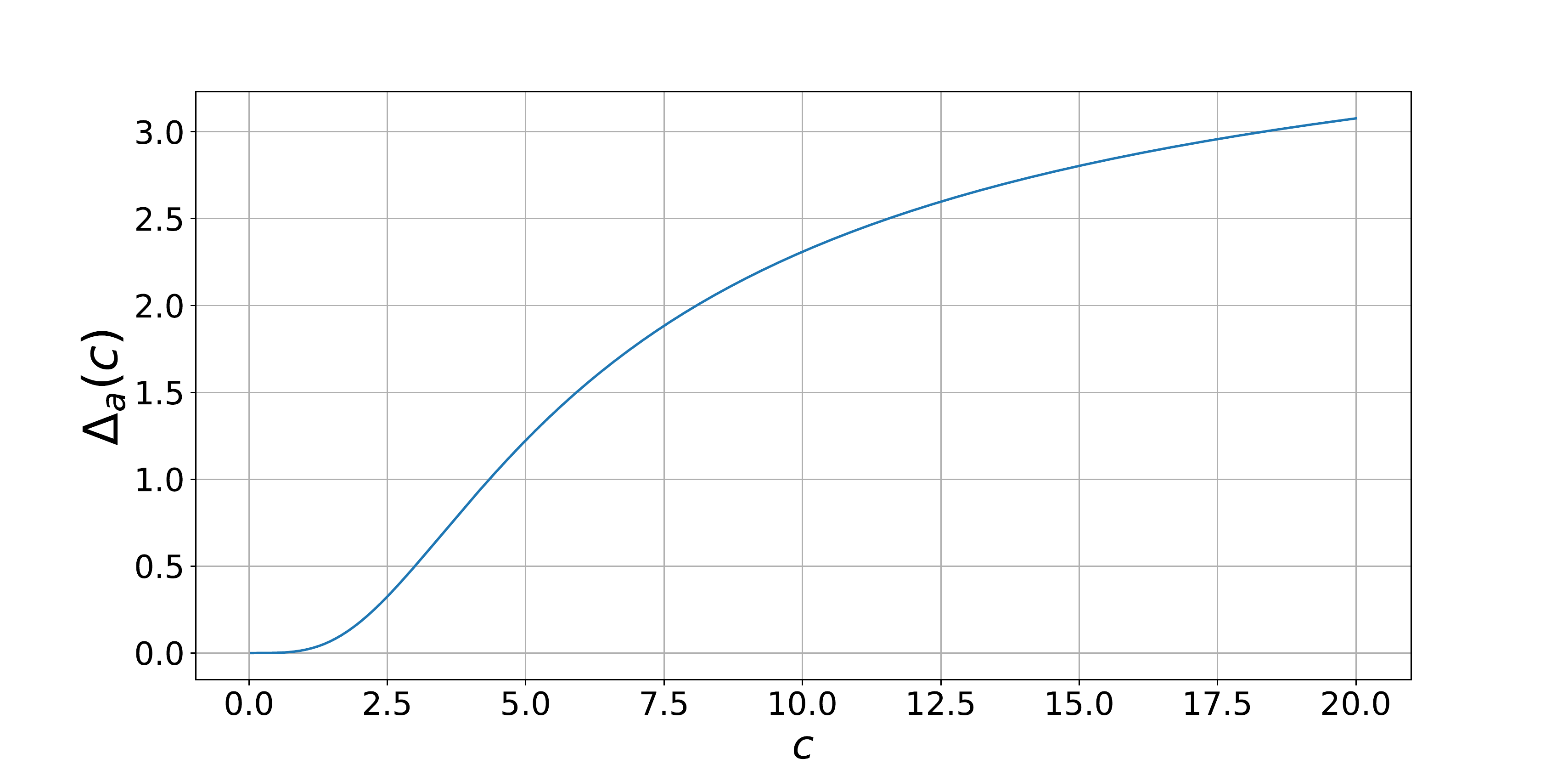} \label{fig:subfigure2}}
        \caption{Scaling functions $\Delta_s(c)$ and $\Delta_a(c)$ for Taylor dispersion in electroosmotic flow }
        \label{fig:figure}
        \label{figosfs}
    \end{figure}

    Here we study Taylor dispersion for neutral particles by a general planar electroosmotic flow\cite{mas06} induced by an electric field applied along a channel composed of two surfaces with different surface charges (see Fig. \ref{schema_channels}(e)). { In the following analysis, we consider point-like particles and are thus considering the case $h=H$.}
    We consider a confined system with  an electrolyte, composed of anions and cations of charge $q$, confined between two infinitely wide parallel plates carrying uniform surface charges or held at constant potential. This problem is thus effectively two dimensional. The  fluid velocity field ${\bf v}$ is incompressible ($\nabla\cdot{\bf v}=0$) and obeys the Stokes equation, 
    \begin{equation}
        \eta  \nabla^2 {\bf v}=\nabla P + \rho_e( E _0 {\bf e}_x -\nabla{\phi}),\label{StokesOsm}
    \end{equation}
    where $\eta$ is the viscosity of the electrolyte and $P$ denotes the pressure. The second term on the right hand side comes from the presence of a charge density $\rho_e$ and an externally applied electric field of magnitude $E_0$ acting along the channel, as well as the electric potential $\phi$ due to the ions themselves. This ionic induced electric potential obeys the Poisson equation
    \begin{equation}
        \epsilon \nabla^2 \phi = -\rho_e,\label{95431}
    \end{equation}
    with $\epsilon$ the dielectric permittivity of the solution. 
    The charge density in the mean field Poisson-Boltzmann approximation is given by
    \begin{equation}
        \rho_e = -2q \rho_b\sinh(q\beta\psi),\label{pb}
    \end{equation}
    with $\rho_b$ the bulk concentration of cations or anions and where $\psi =\phi- E_0 x$ is the total electrostatic potential. If one is interested in the flow at the level of linear response in $E_0$, one can ignore the dependence of $\psi$ on $E_0$ in Eq. (\ref{pb}) to obtain
    \begin{equation}
        \rho_e(z) \simeq -2q \rho_b\sinh(q\beta\phi(z)),\label{pb2}
    \end{equation}
    and thus obtain translational invariance along the channel.With these assumptions, writing Eq. (\ref{StokesOsm}) in the $x$ direction leads to
    \begin{equation}
        \eta  \frac{d^2 v}{dz^2}= \frac{dP}{dx} + \rho_e(z)E_0. 
    \end{equation}
    Assuming that there is no pressure drop along the channel and using Eq. (\ref{95431}), we find
    \begin{equation}
        \eta  \frac{d^2 v}{dz^2}=  \rho_e(z)E_0 = -\epsilon E_0 \frac{d^2\phi(z)}{dz^2},
    \end{equation}
    which can be integrated  to obtain
    \begin{equation}
        v(z) = -\frac{\epsilon E_0}{\eta}[ \phi(z)+ az+b],
    \end{equation}
    where $a$ and $b$ are constants. Imposing no slip boundary conditions at the surfaces $z=\pm H$ then gives
    % \begin{eqnarray}
    % \phi(H)+aH+b&=&0\\
    %\phi(-H)-aH+b&=&0,
    %  \end{eqnarray}
    %  which gives
    %\begin{equation}
    %  b=-\frac{1}{2H}[\phi(H)+\phi(-H)],
    %\end{equation}
    % and
    %\begin{equation}
    %  a= -\frac{1}{2}[\phi(H)-\phi(-H)].
    % \end{equation}
    a velocity field which is completely determined in terms of  the potential $\phi$ and is given by
    \begin{equation}
        v(z) = -\frac{\epsilon E_0}{\eta}\left(\phi(z)-\frac{1}{2H}[\phi(H)-\phi(-H)]z-\frac{1}{2}[\phi(H)+\phi(-H)]\right). 
    \end{equation}
    Note that when $\phi(H)\neq \phi(-H)$, the breaking of symmetry due to a difference in surface potentials generates a linear shear  component to the flow.  In the absence of potential difference, the flow is nearly constant at distances greater than the Debye length from the surface and is referred to as a plug flow. Differing zeta potentials can be generated by using different or treated surfaces \cite{mam13}.
    If we assume that the electrolyte can be treated in the weak coupling regime where $\beta q\phi $ is sufficiently small (and so in particular the surface potential is weak), the Poisson-Boltzmann equation can be linearized giving the Debye-H\"uckel  approximation 
    \begin{equation}
        \frac{d^2 \phi(z)}{dz^2} = \kappa^2 \phi(z),
    \end{equation}
    where 
    \begin{equation}
        \kappa= \sqrt{\frac{2q^2\beta \rho_b}{\epsilon}}
    \end{equation}
    is the inverse Debye screening length. The solution to this equation for this system is then simply
    \begin{equation}
        \phi(z) = \frac{1}{2}\frac{\cosh(\kappa z)}{\cosh(\kappa H)}(\zeta_++\zeta_-)+  \frac{1}{2}\frac{\sinh(\kappa z)}{\sinh(\kappa H)}(\zeta_+-\zeta_-),
    \end{equation}
    where $\zeta_\pm=\phi(\pm H)$ are  the  zeta potentials at the upper and lower surfaces. Physically, the zeta potential often arises due to the presence of a uniform surface charge. It is related to the zeta potential by invoking electro-neutrality which implies that the electric field is zero outside the channel and gives
    \begin{eqnarray}
        -\epsilon\frac{\partial\phi}{\partial z}|_{z=-H}&=& \sigma_-,\\
        \epsilon\frac{\partial\phi}{\partial z}|_{z=H}&=& \sigma_+,
    \end{eqnarray}
    and in terms of the surface charges, we have
    \begin{equation}
        \phi(z) = \frac{1}{2\epsilon \kappa}\frac{\cosh(\kappa z)}{\sinh(\kappa H)}(\sigma_++\sigma_-)+  \frac{1}{2\epsilon\kappa}\frac{\sinh(\kappa z)}{\cosh(\kappa H)}(\sigma_+-\sigma_-),
    \end{equation}
    from which the values of the corresponding zeta potentials can be deduced as a function of surface charges. Working in terms of the zeta potentials, we find  the average drift
    \begin{equation}
        { \left< v\right>_0 = \frac{\epsilon E_0}{2\kappa H \eta}(\zeta_++\zeta_-)\left(\kappa H-\tanh(\kappa H)\right).}
    \end{equation}
    Assuming a homogeneous diffusion constant $D_b$ throughout the system, we find that the diffusion constant due to Taylor dispersion by the flow can be written in the form 
    \begin{equation}
        D_{||}^{(T)}= \frac{\epsilon^2 E^2_0 H^2}{30D_b\eta^2 }\left[ \left(\frac{\zeta_++\zeta_-}{2}\right)^2{\Delta}_s(c) + \left(\frac{\zeta_+-\zeta_-}{2}\right)^2{\Delta}_a(c)\right],
    \end{equation}
    where $H=c/\kappa$ is the height measured in terms of the Debye screening length $\ell_D=1/\kappa$ (so when $c$ is small, the Debye layer is large with respect to the channel length and for $c$ large, the layer is localized near the surface).
    The functions ${\Delta}_s(c)$ (the contribution from the symmetric part of the zeta potentials) and ${\Delta}_a(c)$ (the contribution from the antisymmetric part of the zeta potentials) are given by
    \begin{eqnarray}
        {\Delta}_s(c) &=& \frac{5}{c^4} [12 + 
        2 c^2 - (5c^2+12){\rm sech}^2(c)-9c\tanh(c)] \label{fsf}\\
        {\Delta}_a(c)&=& \frac{1}{c^4}[-60 + 30 c^2 + 4 c^4 + 
        5 c ((3 - 4 c^2) \coth(c) + 9 c\  {\rm cosech}^2(c))]\label{faf}.
    \end{eqnarray}
    This result agrees with that obtained for $\zeta_+=\zeta_-$ in Ref\cite{gri99}. In the limit of small  $c$  (weak screening) and large $c$ (strong screening), we find 
    \begin{eqnarray}
        {\Delta}_s(c) &\simeq& \frac{4c^4}{63} \ {\rm as} \ c\to 0 \ \ \  {\rm and}\   \Delta_s(c)\simeq \frac{10}{c^2} \ {\rm as} \ c\to \infty \\
        {\Delta}_a(c)&\simeq& \frac{4c^4}{189} \ {\rm as} \ c\to 0 \ \ \  {\rm and}\   \Delta_a(c)\simeq 4-\frac{20}{c}\ {\rm as} \ c\to \infty.
    \end{eqnarray}
    The full form of the functions ${\Delta}_s$ and ${\Delta}_a$ are shown in Fig. \ref{figosfs} (a) and (b) respectively.
    
    We thus see that the symmetric contribution to Taylor dispersion is more important for $c$ small, however when $c$ is large, the contribution from the antisymmetric term tends to be bigger. This latter point is due to the fact that the symmetric component is a plug flow which is constant away from the edges and the region where it varies vanishes as $c\to \infty$. In contrast, the anti-symmetric contribution becomes a shear flow in this same limit and thus still disperses the tracer.  We also note that the function ${\Delta}_s(c)$ exhibits a maximum  value  ${\Delta}_s(c_{max})\simeq 0.16$ at $c_{max}\simeq 3.3$, whereas ${\Delta}_a(c)$ increases monotonically as $c$ increases.
    
    The finite time correction is also available within our formalism, we find
    \begin{align}
        C_{||,e}^{(T)}= \frac{E_0^2 \epsilon^2 H^4  }{D_b^2\eta^2 } \left[ \left(\frac{\zeta_++\zeta_-}{2}\right)^2 { \Gamma_s(c)} + \left(\frac{\zeta_+-\zeta_-}{2}\right)^2 { \Gamma_a(c)}\right],
    \end{align}
    with the functions
    \begin{align}
        %g_s(c)&=\frac{-270 - 60 c^2 + 2 c^4 + (270 + 105 c^2 - 2 c^4) \text{sech}^2(c) + 225 \ c \ \text{tanh}(c)}{90 c^6}\\
        { \Gamma_s(c)}& =-\frac{3}{c^6} - \frac{2}{3 c^4} + \frac{1}{45 c^2}  + \frac{5 \tanh(c)}{2 c^5}+ \frac{(270 + 105 c^2 - 2 c^4) \text{sech}^2(c)}{90 c^6} \label{gamma_s}\\
        %g_a(c)&= \frac{1}{c^4} \left(-2 + \frac{2}{c^2} + \frac{c^2}{3} + \frac{17 c^4}{315} + \left(\frac{1}{2c}   + \frac{2 c}{3} - \frac{4 c^3}{15}\right) \text{coth}(c) + \frac{-15 + 2 c^2}{6 \text{sinh}^2 (c)} \right)\\
        { \Gamma_a(c)}&=\frac{2}{c^6} -\frac{2}{c^4}  + \frac{1}{3c^2} + \frac{17 }{315} + \left(\frac{1}{2c^5}   + \frac{2 }{3c^3} - \frac{4 }{15c}\right) \text{coth}(c) + \frac{-15 + 2 c^2}{6 c^4 \text{sinh}^2 (c)} \label{gamma_a}.
    \end{align}
    { that are shown in Fig. \ref{figgamma}. Interestingly they have a similar form to the corresponding scaling functions for the Taylor component of the diffusion constant. While the symmetric term $\Gamma_s$ exhibits a maximum before decaying to zero, the anti-symmetric component $\Gamma_a$  monotonically increases.
     \begin{figure}[h!]
        \centering
        \subfigure[Symmetric function $\Gamma_s(c)$ - Eq. (\ref{gamma_s})]{\includegraphics[width=8cm]{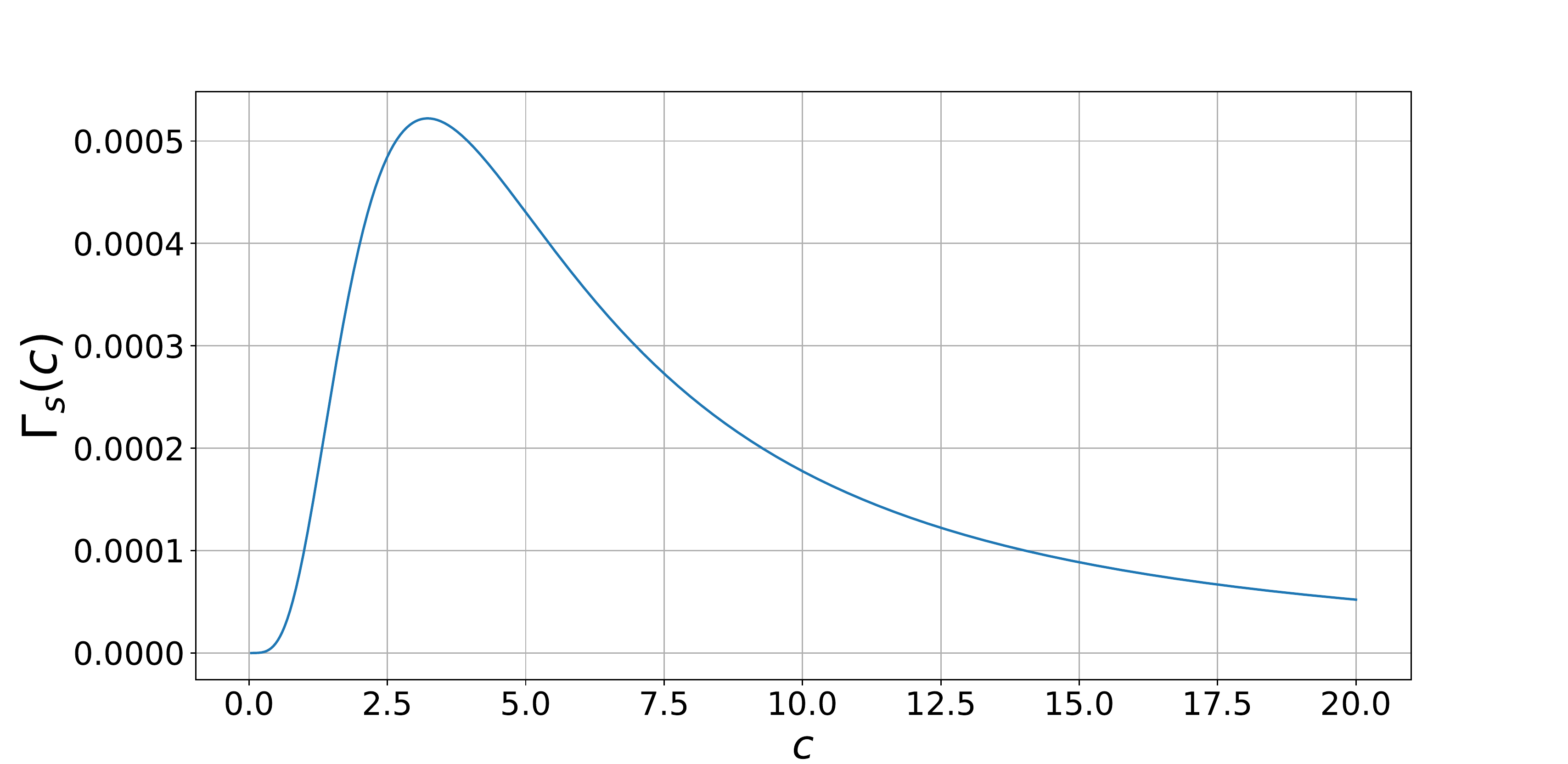}
            \label{fig:gamma_subfigure1}}
        \quad
        \subfigure[Antisymmetric function $\Gamma_a(c)$ - Eq. (\ref{gamma_a})]{ \includegraphics[width=7.5cm]{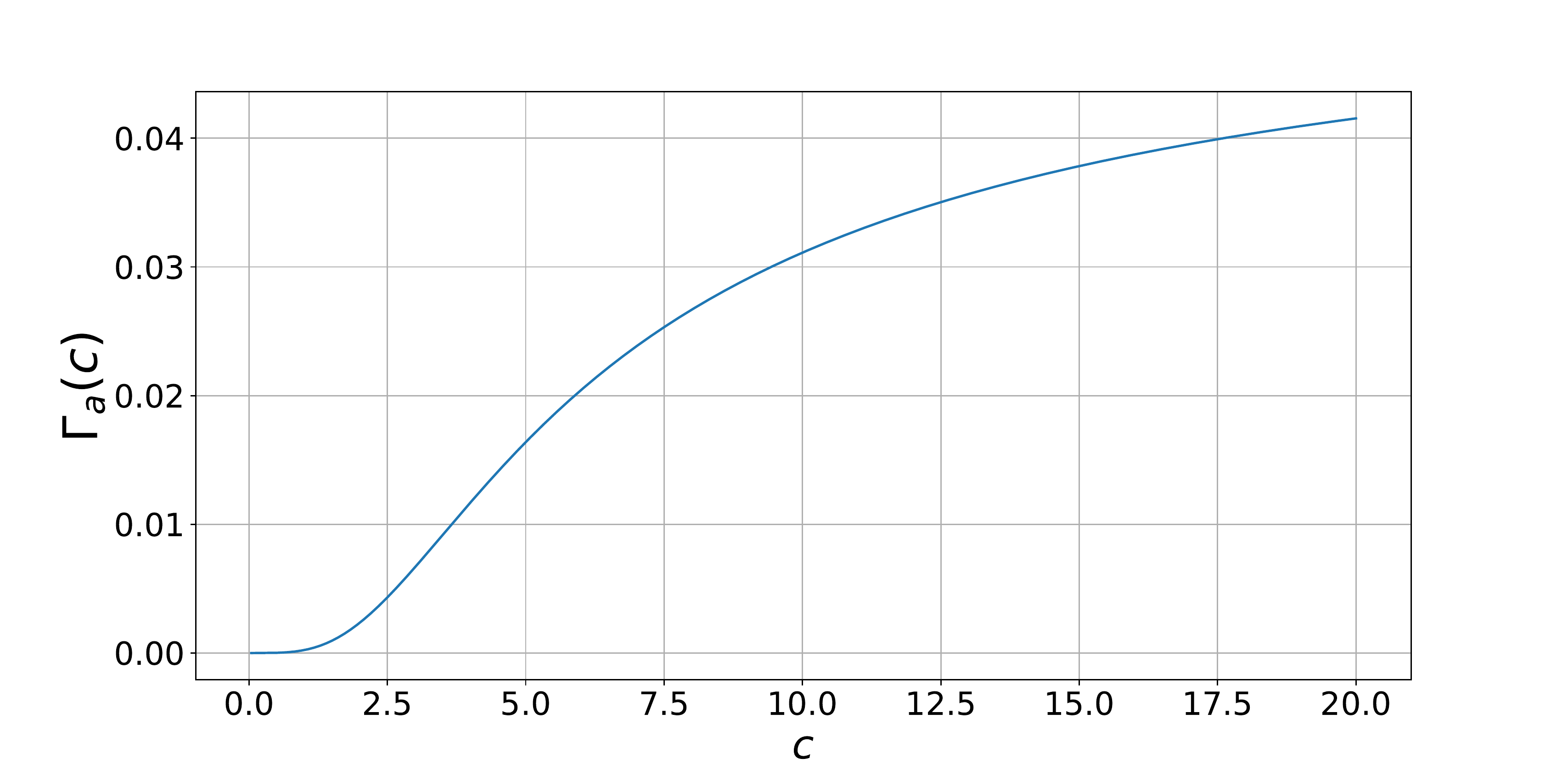} \label{fig:gamma_subfigure2}}
        \caption{{ Scaling functions $\Gamma_s(c)$ and $\Gamma_a(c)$ for the finite time correction for electroosmotic flows}}
        \label{fig:figure}
        \label{figgamma}
    \end{figure}
    
    \noindent We note that using the formalism developed here, one can study the diffusion of a charged particle in 
    an electroosmotic flow. The addition of a tracer charge means that the integrals involved no longer appear to be exactly calculable, however numerically since the computation is straight forward. % In a similar vein, one could also study what happens beyond the Debye-H\"uckel approximation using the full non-linear Poisson-Boltzmann equation \cite{gri20} or even schemes which improve on  mean field theory.
    \section{Conclusion}
    We have revisited the problem of Taylor dispersion in plane and cylindrical channels which are invariant under translation along the channel, that is to say the direction in which the dispersion occurs. For the case of a general diffusion constant $D_\perp$ and potential $V$ acting on the tracer, both of which depend on the distance from the channel center, we have given simple and general formulas for both plane and cylindrical geometries. Essentially, the formulas found encapsulate all known results on dispersion in these channel geometries and further more generalize them. Even results on systems with reaction diffusion with the surfaces can be found and moreover they can be formulated in terms of surface potentials, giving an improved understanding surface mediated effects on Taylor dispersion. Beyond the examples studied here, many more remain to be investigated either analytically or numerically. In particular, the very compact form and generality of the results given here are ideally suited to predict Taylor dispersion in systems where the microscopic parameters such as the diffusion tensor and effective potential can be determined experimentally \cite{fau94,duf01,vil20}.
    
    \section*{Acknowledgements}
    The authors would like to thank Yacine Amarouchene, Vincent Bertin  and Thomas Salez for useful and insightful discussions on this subject.     
    \appendix
    \section{General long  time behavior }\label{agen}
    In the limit $t\rightarrow +\infty$, Eq. (\ref{c43}) becomes
    \begin{equation}
        \left< X_t^2\right>_{cT} \simeq 2 D^{(T)}_{||e} t - 2 C^{(T)}_{||e},\label{latetime}
    \end{equation}
    with 
    \begin{equation}
        D^{(T)}_{||e}=  \int_{-H}^H dz \int_{-H}^H dz' v(z)\sum_{\lambda>0}\frac{\psi_{R\lambda}(z)\psi_{L\lambda}(z')}{\lambda}v(z') p_0(z'),
    \end{equation}
    and 
    \begin{equation}
        C^{(T)}_{||e}=  \int_{-H}^H dz \int_{-H}^H dz' v(z)\sum_{\lambda>0}\frac{\psi_{R\lambda}(z)\psi_{L\lambda}(z')}{\lambda^2}v(z') p_0(z').
    \end{equation}
    We define the pseudo-Green's function by
    \begin{equation}
        G_P(z,z') = \sum_{\lambda>0}\frac{\psi_{R\lambda}(z)\psi_{L\lambda}(z')}{\lambda},
    \end{equation}
    and we note that we can write
    \begin{equation}
        C^{(T)}_{||e}=  \int_{-H}^H dz\ \int_{-H}^H dz'\  v(z)G^2_P(z,z')v(z') p_0(z'),
    \end{equation}
    where we have used the operator notation
    \begin{equation}
        G^2_P(z,z') = \int_{-H}^H dz'' \ G_P(z,z'')G_P(z'',z').
    \end{equation}
    A direct computation then shows that
    \begin{equation}
        \mathcal{H}G_P(z,z')= \sum_{\lambda>0}\psi_{R\lambda}(z)\psi_{L\lambda}(z'),
    \end{equation}
    and using the completeness relation for eigenfunctions  $\sum\limits_{\lambda}\psi_{R\lambda}(z)\psi_{L\lambda}(z')=\delta(z-z')$, we can thus write
    \begin{equation}
        \mathcal{H}G_P(z,z')=\delta(z-z') - p_0(z).
    \end{equation}
    It is also easy to see that $\int_{-H}^H dz \ G_P(z,z') =0$ and that $G_P(z,z')$ obeys the no flux boundary conditions given in 
    Eq. (\ref{noflux}) as a function of $z$. From this, we can also see that 
    \begin{equation}
        f(z)= \int_{-H}^H dz' \ G_P(z,z') v(z')p_0(z'),
    \end{equation}
    obeys 
    \begin{equation}
        \mathcal{H}f(z) = [v(z)-\left< v\right>_0]p_0(z),\label{eqf}
    \end{equation}
    again with the no flux boundary condition Eq. (\ref{noflux}) and the {\em orthogonality} condition 
    \begin{equation}
        \int_{-H}^H dz\  f(z)=0.\label{orth}
    \end{equation}
    This then gives
    the compact formula
    \begin{equation}
        D^{(T)}_{||e}= \int_{-H}^H dz\  v(z) f(z) .\label{dle1}
    \end{equation}
    We can also write 
    \begin{equation}
        D^{(T)}_{||e}=  \int_{-H}^H dz  g(z)v(z) p_0(z),
    \end{equation}
    where 
    \begin{equation}
        g(z) = \int_{-H}^H dz' v(z')G_P(z',z).
    \end{equation}
    Acting on the above equation with $\mathcal{H}^\dagger$ gives
    \begin{equation}
        \mathcal{H}^\dagger g(z) = v(z)-\left< v\right>_0.
    \end{equation}
    The boundary condition on $g(z)$ is $\frac{dg}{dz}|_{z=\pm H}=0$ and $g(z)$ also obeys the integral constraint
    $\int_{-H}^H dz p_0(z) g(z)=0$. However, it is straightforward to see that the solution for $g$ is simply 
    $g(z) = f(z)/p_0(z)$ where $f(z)$ is the solution to Eq. (\ref{eqf}) with the boundary conditions given after Eq. (\ref{eqf}). {This observation} is useful for the computation of the term $C_{||e}$ which can be written as
    \begin{equation}
        C^{(T)}_{||e} = \int_{-H}^H dz dz' dz'' v(z) G_P(z,z'')G_P(z'',z')p_0(z')v(z')= \int_{-H}^H dz \frac{f^2(z)}{p_0(z)}.\label{eqc}
    \end{equation}
    In general, the finite time correction given by the above term can be computed in terms of the  Green's function, it does not require knowledge of  all the eigenfunctions of the problem, and can be used in more general contexts than that considered here \cite{gue15a,gue15b,dea14,dea14b}. We also note that the formula given is easily amenable to numerical analysis and could prove useful in experimental analyses where finite time data can be fitted with the linear  form given in Eq. (\ref{latetime}). We also note from Eq. (\ref{eqc}) that $C^{(T)}_{||e}$ is always positive, this means that a linear fit of the average late time mean squared displacement $y(t) =\left< X_t^2\right>_c$ should cross the $y=0$ axis at positive value of $t$.
    
    We therefore see that the late time correction for Taylor dispersion  can be extracted from the function $f(z)$.
    Finding the solution for $f(z)$ is straightforward, it is given by
    \begin{equation}
        f(z) = \frac{1}{\mathcal{N}}\exp(-\beta V(z))[\left< R\right>_0 - R(z)],
    \end{equation}
    where $R(z)$ is given by
    \begin{equation}
        R(z)= \int_{-H}^z dz'\ \frac{J(z')\exp(\beta V(z'))}{D_{\perp}(z')},
    \end{equation}
    and $J(z)$ is given by
    \begin{equation}
        J(z) = \int_{-H}^z dz' \exp(-\beta V(z'))[v(z)-\left< v\right>_0].
    \end{equation}
    The expression for $D^{(T)}_{||e}$ can be rewritten in a more symmetric fashion, using the Eq. (\ref{dle1}) and the condition Eq. (\ref{orth}) we have 
    \begin{equation}
        D^{(T)}_{||e}= \int_{-H}^H dz\  (v(z)-\left< v\right>_0) f(z) = \frac{1}{\mathcal{N}}\int_{-H}^H dz \frac{dJ(z)}{dz}(\left< R\right>_0- R(z)).
    \end{equation}
    \noindent Integrating by parts and using the no-flux boundary condition $J(\pm H)=0$ then yields
    \begin{equation}
        D^{(T)}_{||e} = \frac{1}{\mathcal{N}}\int_{-H}^H dz \ \frac{J^2(z)\exp(\beta V(z))}{D_\perp(z)} .\label{dtaylor}
    \end{equation}
    In addition, using the above expressions, the late time correction can be rewritten 
    \begin{equation}
        C^{(T)}_{||e} = \left< R^2\right>_0- \left< R\right>^2_0,
    \end{equation}
    which is of course equivalent to Eq. (\ref{eqc}).
    
    \section{Derivation of the dispersion for strong wall interactions using saddle-point approximation}
    \label{AppendixWall}
    Let us consider a channel where the potential is localized near the wall.
    Two cases spring to mind, one where the potential has a single minimum at a position
    $z_{min}$ close to the wall and no energy barrier to attain it, secondly one could imagine that there is an energy barrier due to a maximum in the potential at $z_{max}$ to the right of $z_{min}$ at the left wall.
    
    For systems possessing a  symmetry under $z\to-z$, we can write the Taylor dispersion contribution to the diffusion constant along the channel as 
    \begin{equation}
        D_{||e}^{(T)} = \frac{2}{\mathcal{N}}\int_{-H+\ell}^0 dz \frac{J^2(z)}{D_\perp(z)} + \frac{2}{\mathcal{N}}\int_{-H}^{-H+\ell} dz \exp(\beta V(z))\frac{J^2(z)}{D_\perp(z)}.
    \end{equation}
    In the region $[-H,-H+\ell]$ we will treat the problem via the saddle point approximation (keeping the fluctuation term) and  assuming that the velocity field $v(z)$ and diffusion constants vary much more slowly than the potential $V(z)$ . 
    For a potential with a minimum near the wall but with a maximum between the wall and the bulk, we find that
    the partition function is given by
    \begin{eqnarray}
        \mathcal{N} = 2\left[H- \ell + \frac{\sqrt{2\pi}}{\sqrt{|\beta V''(z_{min})|}}\exp(-\beta V(z_{min}))\right].
    \end{eqnarray}
    From the above, we see that the equilibrium probability density in the bulk can be written as
    \begin{equation}
        p_{b0} = \frac{1}{2\left[H- \ell+ \frac{\sqrt{2\pi}}{\sqrt{|\beta V''(z_{min})|}}\exp(-\beta V(z_{min})\right]},
    \end{equation}
    while the probability to be on either of the surfaces is
    \begin{equation}
        P_{s0} = \frac{\frac{\sqrt{2\pi}}{\sqrt{|\beta V''(z_{min})|}}\exp(-\beta V(z_{min}))}{2\left[H- \ell + \frac{\sqrt{2\pi}}{\sqrt{|\beta V''(z_{min})|}}\exp(-\beta V(z_{min}))\right]},
    \end{equation}
    the saddle point approximation respects the conservation of probability as we see that
    $2P_{s0}+ 2p_{b0}(H-\ell)=1$.
    Clearly, to be in a limit where the surface affects the dispersion, one needs to consider the limit where $P_{s0}$ is 
    ${\cal O}(1)$. 
    Carrying out a similar analysis, we find
    \begin{eqnarray}
        D_{||e}^{(T)}&\simeq& 2 p_{b0}\int_{-H+\ell}^{0} dz \frac{J^2(z)}{D_\perp(z)}+ 2p_{b0}\frac{\sqrt{2\pi}}{\sqrt{|\beta V''(z_{max})|}}\frac{J^2(z_{max})}{D_\perp(z_{max})}\exp(\beta V(z_{max}))\nonumber \\
        &=& D_{||e}^{(Tb)} + D_{||e}^{(Ts)},
    \end{eqnarray}
    where in the last line above, we have decomposed the advective contribution to the diffusion constant into a
    bulk $D_{||e}^{(Tb)}$ and a surface term $D_{||e}^{(Ts)}$which is sensitive to the maximum in the potential (if present). The surface contribution can be written as 
    \begin{equation}
        D_{||e}^{(Ts)}= 2\frac{p_{b0}^2}{P_{s0}} J^2(z_{max})\tau,
    \end{equation}
    where
    \begin{equation}
        \tau = \frac{2\pi}{\sqrt{|\beta^2 V''(z_{max})V''(z_{min})|}D_\perp(z_{max})}\exp(\beta [V(z_{max})-V(z_{min})])
    \end{equation}
    can be shown to be the time to cross the energy barrier between $z_{min}$ and $z_{\max}$ assuming that the diffusion constant can be treated as a constant in this region (in particular $D_\perp(z_{max})\simeq D_\perp(z_{min})$). In the absence of a maximum of the potential before the wall, one can take $\tau=0$. We note that this term is always positive and so any barrier in potential between the surface and the bulk will in general increase Taylor dispersion.
    
    In the left edge region where  $z<-H+\ell$, we find that the current is given by
    \begin{equation}
        J(z) \simeq 0 \ {\rm for}\ z< z_{min}
    \end{equation}
    and 
    \begin{equation}
        J(z) \simeq \frac{P_{s0}}{p_{b0}}(v(z_{min})-\left< v\right>_0)\  {\rm for}\ z> z_{min} ,\label{jedge}
    \end{equation}
    while, in the bulk, for $z>-H+\ell$ 
    \begin{equation}
        J(z) =\frac{P_{s0}}{p_{b0}}(v(z_{min})-\left< v\right>_0) + \int_{-H+\ell}^z dz' (v(z')-\left< v\right>_0).
    \end{equation}
    We also have
    \begin{equation}
        \left< v\right>_0= 2 P_{s0} v(z_{min}) + 2 p_{b0} \int_{-H+\ell}^0 dz \ v(z)\label{vo},
    \end{equation}
    and which can also be written as 
    \begin{equation}
        \left< v\right>_0 = \left< v\right>_b+ 2 P_{s0} (v(z_{min})-\left< v\right>_b),
    \end{equation}
    where 
    \begin{equation}
        \left< v\right>_b= \frac{1}{H-\ell}\int_{-H-\ell}^0dz\ v(z)
    \end{equation}
    is the average of the advection velocity field in the bulk region of the channel  $[-H+\ell,H-\ell]$ where there is no potential.
    From Eq. (\ref{vo}) and Eq. (\ref{jedge}), we find
    \begin{equation}
        J(z_{max})= \frac{P_{s0}}{p_{b0}}(v(z_{min})-\left< v\right>_0),
    \end{equation}
    which yields 
    \begin{equation}
        \begin{split}
            D_{||e}^{(Ts)}&= 2P_{s0} \tau (v(z_{min})-\left< v\right>_0)^2\\
            &= 2P_{s0} \tau(1-2P_{s0})^2(v(z_{min})-\left< v\right>_b)^2.
        \end{split}
    \end{equation}
    Note that this term is zero when $P_{s0}=0$, so there is zero probability to be in the wall region, but also when $\tau=0$, that is to say it requires the existence of the local maximum in the potential before the local minimum next to the wall. 
    For the {\it bulk term}, we find
    \begin{eqnarray}
        D_{||e}^{(Tb)}&=\frac{2}{p_{b0}} \int_{-H+\ell}^{0} dz \frac{\left(P_{s0}(v(z_{min})-\left< v\right>_0) +p_{b0} \int_{-H+\ell}^z dz' (v(z')-\left< v\right>_0)\right)^2}{D_\perp(z)} \nonumber \\
        &= 2 p_{b0}\int_{-H+\ell}^{0} dz \frac{\left(-2zP_{s0}(v(z_{min})-\left< v\right>_b) + \int_{-H+\ell}^z dz' (v(z')-\left< v\right>_b)\right)^2}{D_\perp(z)}.
    \end{eqnarray} 
    Expanding the integrand yields
    \begin{eqnarray}
        D_{||e}^{(Tb)}&=& 8 p_{b0}P_{s0}^2(v(z_{min})-\left< v\right>_b)^2\int_{-H+\ell}^0 dz\frac{z^2}{D_\perp(z)}+
        2 p_{b0}\int_{-H+\ell}^{0} dz \frac{\left(\int_{-H+\ell}^z dz' (v(z')-\left< v\right>_b)\right)^2}{D_\perp(z)}\nonumber \\
        &+& 8p_{b0}P_{s0}(\left< v\right>_b-v(z_{min}))\int_{-H+\ell}^0 dz\frac{z\int_{-H+\ell}^z dz' (v(z')-\left< v\right>_b)}{D_\perp(z)},
    \end{eqnarray}
    and we note that the second term can be written as 
    \begin{equation}
        2 p_{b0}\int_{-H+\ell}^{0} dz \frac{\left(\int_{-H+\ell}^z dz' (v(z')-\left< v\right>_b)\right)^2}{D_\perp(z)}=
        (1-2P_{s0})D_{||e}^{(Tb)}(H-\ell,V=0),
    \end{equation}
    where $D_{||e}^{(T)}(H-\ell,V=0)$ is the Taylor dispersion generated part of the diffusion constant for the system with size $H-\ell$ and without any potential, that is to say the same $v(z)$ and $D_{\perp}(z)$ but restricted to the region $[-H+\ell, H-\ell]$.
    In the case where $D_\perp(z)$ can be approximated by a constant in the bulk region, so assuming that modifications of the diffusion constant, due to the finite particle size, only play a role in the surface layer, we find
    \begin{eqnarray}
        D_{||e}^{(Tb)}&=& 8 p_{b0}P_{s0}^2(v(z_{min})-\left< v\right>_b)^2\frac{(H-\ell)^3}{3D_\perp}+
        2 \frac{p_{b0}}{D_\perp}\int_{-H+\ell}^{0} dz \left(\int_{-H+\ell}^z dz' (v(z')-\left< v\right>_b)\right)^2
        \nonumber \\
        &-& \frac{8p_{b0}P_{s0}(v(z_{min})-\left< v\right>_b)}{D_\perp}\int_{-H+\ell}^0 dz\ {z\int_{-H+\ell}^z dz' (v(z')-\left< v\right>_b)}.
    \end{eqnarray}
    Now integrating the last term by parts yields
    \begin{eqnarray}
        D_{||e}^{(Tb)}&=& 8 p_{b0}P_{s0}^2(v(z_{min})-\left< v\right>_b)^2\frac{(H-\ell)^3}{3D_\perp}+
        2 \frac{p_{b0}}{D_\perp}\int_{-H+\ell}^{0} dz \left(\int_{-H+\ell}^z dz' (v(z')-\left< v\right>_b)\right)^2
        \nonumber \\
        &+& \frac{4p_{b0}P_{s0}(v(z_{min})-\left< v\right>_b)}{D_\perp}\int_{-H+\ell}^0 dz\ {z^2 (v(z)-\left< v\right>_b)}.
    \end{eqnarray}
    
    We now compare with the results of a surface reaction diffusion model \cite{bre93,lev12}. In these previous studies a Poiseuille flow with no-slip boundary conditions at the boundary was used so we set
    \begin{equation}
        v(z) = v_0\left(1-\frac{z^2}{H_s^2}\right),
    \end{equation}
    where $H_s=h$ corresponds to the no slip boundary condition and $H_s>h$ corresponds to a finite slip length.
    We now  have a surface layer with $\ell=0$ and so we take $z_{min}=-H$. In addition, no variation in the diffusion constant perpendicular to the channel was considered so we take $D_\perp(z) = D_b$ where $D_b$ is the bulk diffusion constant. Finally, we assume that the ratio of the probability, at equilibrium,  to be on the surface to that to be in the bulk
    \begin{equation}
        \alpha = \frac{\frac{\sqrt{2\pi}}{\sqrt{|\beta V''(z_{min})|}}\exp(-\beta V(z_{min})])}{H},
    \end{equation}
    is of order one. This definition now implies that
    \begin{equation}
        P_{s0}= \frac{1}{2}\frac{\alpha}{1+\alpha}; \ p_{b0}= \frac{1}{2H}\frac{1}{1+\alpha}, \label{rap}
    \end{equation}
    and we thus obtain 
    \begin{eqnarray}
        \frac{D_{||e}^{(Tb)}}{D_b}= Pe^2 \frac{2 + 18\alpha + 51\alpha^2}{105(1+\alpha)^3},
    \end{eqnarray}
    where 
    \begin{equation}
        Pe=\frac{2v_0H^3}{3 H_s^2 D_b},\label{pe}
    \end{equation}
    is the relevant P\'eclet number. The surface contribution to Taylor dispersion yields
    \begin{equation}
        \frac{D_{||e}^{(Tb)}}{D_b}= \frac{4H^4}{9D_bH_s^4}v_0^2\frac{\alpha \tau}{(1+\alpha)^3}= Pe^2\frac{\alpha }{(1+\alpha)^3}\frac{D_b\tau}{H^2}.
    \end{equation}
    If we assume that the diffusion constant on the surface region is given by $D_{||s}$ and $D_{\perp s}$ (along and perpendicular to the surface), we obtain the result
    \begin{equation}
        \frac{D_{\parallel e}}{D_b} = \frac{1+\alpha s_{\parallel}}{1+\alpha} + Pe^2\left(\frac{2 + 18\alpha + 51\alpha^2}{105(1+\alpha)^3}+ \frac{\alpha }{(1+\alpha)^3}\frac{D_b\tau}{H^2}\right),\label{sgen}
    \end{equation} 
    where $s_{\parallel}= D_{s \parallel}/D_b$ in the first term, and we note that this first term simply corresponds to the change in molecular diffusion along the channel due to the interaction with the wall. In this particular case, we have 
    \begin{equation}
        \tau = \frac{2\pi}{\beta\sqrt{| V''(z_{max})V''(z_{min})|}D_{s\perp }}\exp(\beta [V(z_{max})-V(z_{min})]),
    \end{equation}
    and we also note that the term $\tau_T= H^2/D_b$ in the last term of Eq. (\ref{sgen}) is commonly referred to  as the Taylor time, the time to cross the bulk of the system from one side to another or equivalently the equilibration time for reflected Brownian motion in the channel. 
    
 \noindent {\bf Data availability statement}: Data sharing is not applicable to this article as no new data were created or analyzed in this study.
    
\end{document}